\documentclass[apj]{emulateapj}

\usepackage{natbib}

\usepackage{graphicx}
\usepackage[percent]{overpic}
\usepackage{lipsum}
\usepackage{multirow}
\usepackage{amssymb}
\usepackage{color}
\usepackage{rotating}
\usepackage{mwe,tikz}
\usepackage[percent]{overpic}
\usepackage{mathtools}
\usepackage{physics}
\usepackage{amsmath}
\usepackage[utf8x]{inputenc}
\usepackage{hyperref}
\usepackage{wasysym}
\usepackage{soul}

\newcommand{\msun}{${\rm M_{\sun}}$}
\def\ltsima{$\; \buildrel < \over \sim \;$}
\def\simlt{\lower.5ex\hbox{\ltsima}}
\def\gtsima{$\; \buildrel > \over \sim \;$}
\def\simgt{\lower.5ex\hbox{\gtsima}}
\def\kms{{\rm\,km\,s^{-1}}}
\def\kmmss{{\rm\,km^2\,s^{-2}}}
\def\pc{{\rm\,pc}}
\def\kpc{{\rm\,kpc}}

\def\msun{{\rm\,M_\odot}}


\def\deg{^\circ}

\def\Gyr{{\rm\,Gyr}}

\def\masyr{{\rm\,mas \, yr^{-1}}}

\def\ltsima{$\; \buildrel < \over \sim \;$}
\def\gtsima{$\; \buildrel > \over \sim \;$}

\def\Gaia{{\it Gaia\,}}
\defcitealias{Ibata_GaiaEDR3}{I20}

\slugcomment{Accepted in The Astrophysical Journal}

\shorttitle{Understanding the nature of the LMS-1 stellar stream}
\shortauthors{Malhan et al.}

\begin{document}

\title{Evidence of a dwarf galaxy stream populating the inner Milky Way Halo}

\email{Corresponding author:kmalhan07@gmail.com}

\author{Khyati Malhan\altaffilmark{1}, Zhen Yuan\altaffilmark{2}, Rodrigo A. Ibata\altaffilmark{2}, Anke Arentsen\altaffilmark{2}, Michele Bellazzini\altaffilmark{3}, Nicolas F. Martin\altaffilmark{2,4}}

\altaffiltext{1}{The Oskar Klein Centre, Department of Physics, Stockholm University, AlbaNova, SE-10691 Stockholm, Sweden}
\altaffiltext{2}{Universit\'e de Strasbourg, CNRS, Observatoire Astronomique de Strasbourg, UMR 7550, F-67000 Strasbourg, France}
\altaffiltext{3}{INAF  -  Osservatorio  di  Astrofisica  e  Scienza  dello  Spazio,via Gobetti 93/3, 40129 Bologna, Italy}
\altaffiltext{4}{Max-Planck-Institut f\"ur Astronomie, K\"onigstuhl 17, D-69117, Heidelberg, Germany}

\begin{abstract}

Stellar streams produced from dwarf galaxies provide direct evidence of the hierarchical formation of the Milky Way. Here, we present the first comprehensive study of the ``LMS-1'' stellar stream, that we detect by searching for wide streams in the {\it Gaia} EDR3 dataset using the \texttt{STREAMFINDER} algorithm. This stream was recently discovered by Yuan et al. (2020). We detect LMS-1 as a $60\deg$ long stream to the north of the Galactic bulge, at a distance of $\sim 20$ kpc from the Sun, together with additional components that suggest that the overall stream is completely wrapped around the inner Galaxy. Using spectroscopic measurements from LAMOST, SDSS and APOGEE, we infer that the stream is very metal poor (${\rm \langle [Fe/H]\rangle =-2.1}$) with a significant metallicity dispersion ($\sigma_{\rm [Fe/H]}=0.4$), and it possesses a large radial velocity dispersion (${\rm \sigma_v=20 \pm 4\,km\,s^{-1}}$). These estimates together imply that LMS-1 is a dwarf galaxy stream. The orbit of LMS-1 is close to polar, with an inclination of $75\deg$ to the Galactic plane. Both the orbit and metallicity of LMS-1 are remarkably similar to the globular clusters NGC~5053, NGC~5024 and the stellar stream ``Indus''. These findings make LMS-1 an important contributor to the stellar population of the inner Milky Way halo. 

\end{abstract}

\section{Introduction}\label{sec:Introduction}

The standard $\Lambda$ cold dark matter ($\Lambda$CDM) cosmological model predicts that the Milky Way halo formed hierarchically by the accretion of numerous progenitor galaxies (e.g., \citealt{Bullock2005, Pillepich2018, Kruijssen_2019}). These progenitor galaxies initially brought in their own stellar populations (in the form of stars and globular clusters). Over time, the progenitor galaxies were disrupted by the tidal forces of the Galactic potential, eventually leading to the dispersal of their stars. Today, these accreted stellar populations can be observed as a part of our Galactic halo. Since these accretion events are expected to have contributed significantly to the Milky Way's stellar halo (cf. \citealt{Chiba2000, Bell2008, Nissen2010, Belokurov2018, Helmi2018, Myeong2019, Matsuno2019, Koppelman2019, Yuan2020a, Naidu2020}), it is important to identify and study these events in order to understand the formation history of our Galaxy. For instance, the number of dwarf galaxy streams observed in the Milky Way can effectively place a lower limit on the number of accretions that have taken place into our Galaxy.

Low mass satellites that are not on extremely radial orbits can give rise to stellar streams as they disrupt. These structures trace approximately the orbit of the progenitor in the Galactic potential. Therefore, mapping the positions and velocities of stream stars provides a first estimate of the orbital path along which the accretion took place and along which the progenitor deposited its stellar contents. This orbital property of streams also makes them extremely powerful probes of the dark matter distribution of the Milky Way \citep{Law2010}. In this regard, particularly interesting are the `polar' streams\footnote{By `polar streams', we refer to those streams whose orbital planes are inclined almost perpendicularly to the Galactic plane.} since they are very useful for gauging the flattening of the dark matter halo \citep{Erkal_2016}. These important goals of studying the dark matter and the assembly history of our Galaxy has motivated much of the work in detecting and analysing dwarf galaxy streams.

Most of the dwarf galaxy streams (e.g., Sagittarius, \citealt{Majewski2003}, Orphan, \citealt{Grillmair_Orphan2006, Belokurov2007}, Cetus , \citealt{Newberg_Cetus2009}, Tucana III, \citealt{Li_TucIII_stream2018}) appear to be dynamically-young systems. That is, these systems were perhaps only recently accreted into the Milky Way ($\simlt 3-6\Gyr$ ago), long after our Galaxy reached its current size and mass. Such recent accretion times are suggested by the studies that have attempted to construct dynamical models of these streams (e.g., see \citealt{Erkal_2018_TucIII} for Tucana III, \citealt{Erkal2019_Orphan} for Orphan). For Sagittarius, this time frame is based on the starburst episode in the disk $\sim 6\Gyr$ ago, that has been linked with the first pericentric passage of this massive dwarf galaxy \citep{Ruiz-Lara2020, Lian2020}. Furthermore, many of these streams are highly coherent in phase-space, implying that they have not yet had enough time to phase mix in the halo --- again favouring a recent accretion scenario. However, the hierarchical paradigm of galaxy formation suggests that several dwarf galaxies must have accreted into the Milky Way at earlier times (e.g., $\simgt 8-10\Gyr$ ago) when our Galaxy was smaller in size and still rapidly forming (e.g., \citealt{Zhao_2009}). If this scenario is true, then it raises the question: at the present day, does the Milky Way contain signatures of dynamically-old dwarf galaxy streams, or are they completely phase-mixed and impossible to detect?

If the Galaxy grew ``inside-out'', these dynamically-old streams are likely to be discovered in the inner regions of the halo ($\simlt 20-30\kpc$), where the progenitor galaxy would have originally been accreted and disrupted. We expect it to be challenging to detect these ancient dwarf galaxy fragments amidst the other populations of the stellar halo, primarily because of their low contrast after undergoing significant phase-mixing. However, the excellent astrometric dataset from the ESA/\Gaia mission \citep{Prusti_2016} may provide the key to their detection.

Here, we study the properties of the so-called ``LMS-1'' stream. This stream was discovered by \cite{Yuan2020} using RR Lyrae and blue horizontal-branch stars in \Gaia DR2 \citep{GaiaDR2_2018_Brown, Lindegren2018}. \cite{Yuan2020} designated this structure as the ``low-mass stream'' (LMS-1 for short) to be consistent with their inference that the progenitor of this stream was likely a low-mass dwarf galaxy; given its low mean metallicity ($\langle {\rm [Fe/H]}\rangle \sim -2$) and broad physical width. LMS-1 is the same halo structure referred to as ``Wukong'' in \cite{Naidu2020}, who independently discovered it using the combination of \Gaia and H3 Spectroscopic Survey \citep{Conroy_H3_2019}. No progenitor of LMS-1 has so far been detected, likely because it has completely disrupted. \cite{Yuan2020} originally discovered LMS-1 as a stream located towards the north of the Galactic bulge, and a rough orbital analysis indicated that it possesses a very small apocentre of $\sim20\kpc$ and a pericentre of $\sim 10\kpc$. As far as we are aware, no other dwarf galaxy stream, or even an intact dwarf galaxy, is known to orbit the Milky Way at such small orbital radius\footnote{Indeed, among the $46$ dwarf galaxies analysed by \cite{Li_MWdwarfs_GaiaEDR3}, the smallest orbital radius is that of Tuc~III with apocentre $\sim 46\kpc$ and a pericentre $\sim 3\kpc$.}. This makes LMS-1 an extremely interesting system, and the motivation of our current study is to answer the following questions. What is the nature of LMS-1? How are LMS-1 stars distributed in the Milky Way? Which other objects of the Milky Way (e.g., globular clusters, other streams) were accreted inside the LMS-1's parent galaxy? How do the overall properties of LMS-1 compare with other dwarf galaxy streams of the Milky Way? 

To answer these questions about the LMS-1 stream, we re-analyse this system using the superb new astrometric data from \Gaia EDR3 \citep{GaiaEDR3_Brown_2020}. This paper is arranged as follows. Section~\ref{sec:Data} describes the data used and briefly summarises the detection algorithm and adopted parameters. Section~\ref{sec:orbit} explains the method for constraining the LMS-1's orbit. In Section~\ref{sec:vel_disp} we estimate the velocity dispersion of the LMS-1 stream. Section~\ref{sec:RRLyare_sel} details the procedure of selecting the RR Lyrae candidates of LMS-1. In Section~\ref{sec:Nbody_model} we construct and study dynamical models of LMS-1. In Section~\ref{sec:ActionEnergy} we search for companion globular clusters and other streams that move on similar orbits as that of LMS-1 and were possibly accreted into the Milky Way along with LMS-1. Finally, we discuss our findings and conclude in Section~\ref{sec:Discussion_Conclusion}.

\begin{figure*}
\begin{center}
\vspace{-0.3cm}
\includegraphics[width=0.96\hsize]{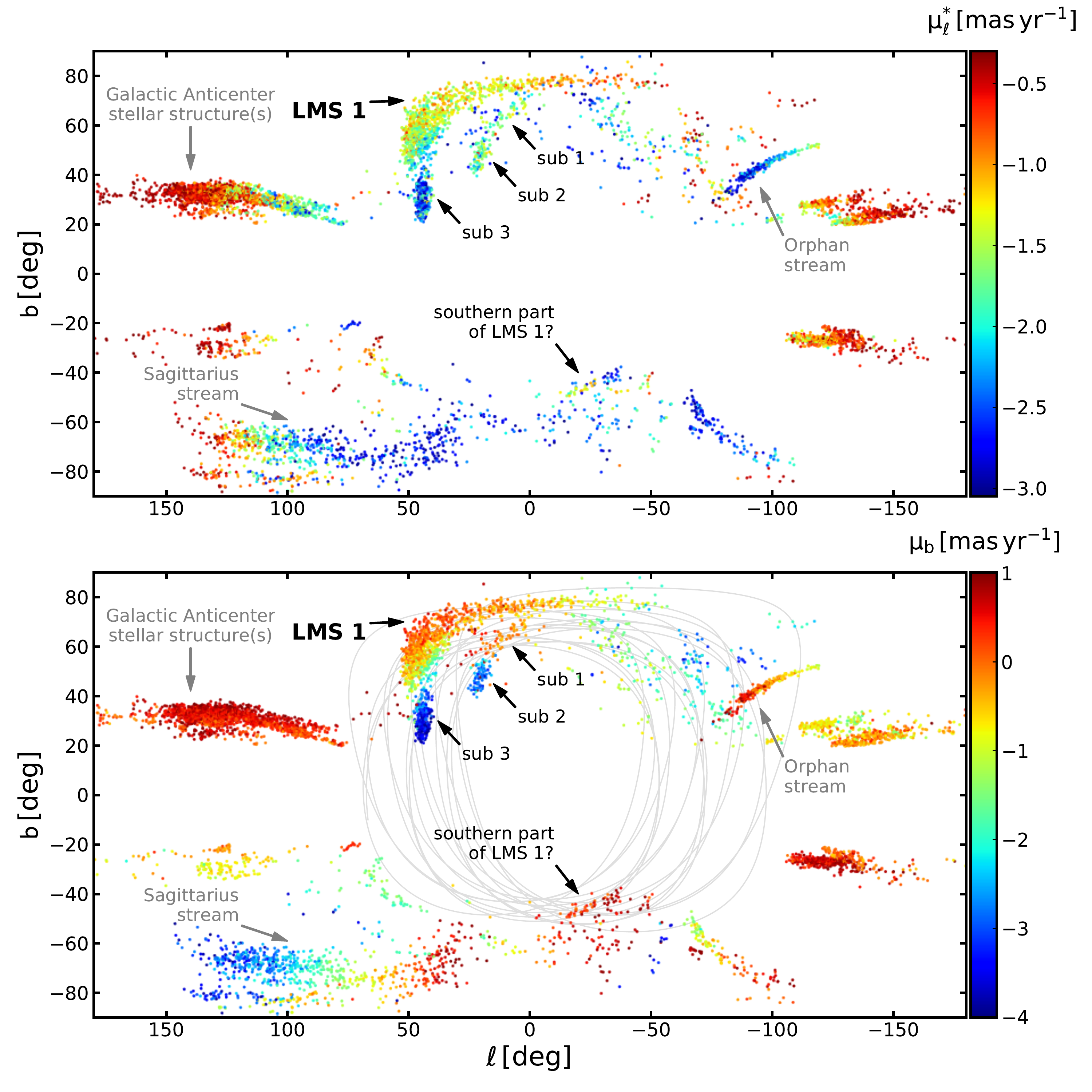}
\end{center}
\vspace{-0.5cm}
\caption{The LMS-1 stream detected in \Gaia EDR3 using the \texttt{STREAMFINDER} algorithm. The Galactic sky maps show stars that are identified as stream-like with $>10\sigma$ confidence upon using a stream template of Gaussian width $0.5\kpc$ and searching for streams in the distance interval $10< d_{\odot}<30\kpc$ (assuming a metal-poor stellar population model with Age=$12.5\Gyr$ and [Fe/H]=$−2.2$). Also, only those stars are shown that possess proper motions in the range as highlighted by the color bars. The top and bottom panels show the distribution of proper motion in $\mu^{*}_{\ell}$ and $\mu_b$, respectively. In addition to detecting the LMS-1 stream, some of the previously known broad stellar streams are also detected (e.g., ``Orphan'', ``Galactic Anticenter stellar structure(s)'' and parts of ``Sagittarius''). We also discover some of the off-stream sub-structures that are likely associated with LMS-1 (labelled as sub-1, sub-2 and sub-3). Finally, the agglomeration of stars at $(\ell,b)\sim(-40\deg,-60\deg)$ is probably the continuation of LMS-1 in the southern galactic sky. In the bottom panel, the gray curve highlights the orbit of LMS-1 derived in Section~\ref{sec:orbit}.}
\label{fig:Fig_SF_map}
\end{figure*}
\section{Data and LMS-1 detection}\label{sec:Data}

To detect LMS-1, we process the \Gaia EDR3 catalogue using the \texttt{STREAMFINDER} algorithm \citep{Malhan_SF_2018, Malhan_PS1_2018}. The application of \texttt{STREAMFINDER} on \Gaia DR2 and EDR3 has been previously detailed in \cite{Malhan_Ghostly_2018, Ibata_norse_2019, Malhan_Kshir2019} and \citet[hereafter \citetalias{Ibata_GaiaEDR3}]{Ibata_GaiaEDR3}. In brief, \texttt{STREAMFINDER} is effectively a friend-finding algorithm that considers each star in a dataset in turn, and searches for similar stars along all possible orbits of the star under consideration. To search for streams this way, \texttt{STREAMFINDER} analyses the following \Gaia measurements of stars: positions ($\alpha,\delta$), parallaxes ($\overline{\mathbb{\omega}}$), proper motions ($\mu^{*}_{\alpha}, \mu_{\delta}$) and photometry (in the $G,G_{\rm BP},G_{\rm RP}$ pass-bands). For this, we first correct the \Gaia magnitudes for dust extinction using the \cite{Schlegel1998} maps, adopting the re-calibration by \cite{Schlafly2011}, assuming the extinction ratios $A_{\rm G}/A_{\rm V}=0.86117, A_{\rm BP}/A_{\rm V}= 1.06126, A_{\rm RP}/A_{\rm V}= 0.64753$ (as listed on the web interface of the PARSEC isochrones, \citealt{Bressan2012}). Henceforth, all magnitudes and colors refer to these extinction-corrected values. We only consider stars down to a limiting magnitude of $G_0=20$ and discard the fainter sources to minimise the spatial inhomogeneities in the maps. Moreover, \texttt{STREAMFINDER} integrates orbits to find streams, and for this we adopt the Galactic potential model 1 of \cite{DehnenBin1998}. Furthermore, we adopt the Sun's Galactocentric distance from \cite{Gravity2018} and the Sun's galactic velocity from \cite{Reid2014_Sun, Schornich2010_Sun}. These parameters are required to transform the measured Heliocentric positions and velocities of stars in the Galactocentric coordinates to integrate orbits, and also to transform the orbits back in the Heliocentric frame for the comparison with the measurements. 

Our overall procedure for detecting streams is identical to the one employed in \citetalias{Ibata_GaiaEDR3}. However, we made changes in some of the parameters so as to specifically find broad and metal-poor streams (as expected from dwarf galaxies). In particular, we adopt a stream width of (Gaussian) dispersion $0.5\kpc$, and allowed the algorithm to search for ``friends'' along $20\deg$-long orbits. Also, we ran the algorithm using a specific stellar population template from the PARSEC library of age and metallicity (Age, [Fe/H])=($12.5\Gyr,-2.2$). Moreover, the algorithm was only allowed to search for distance solutions in the Heliocentric distance range $d_{\odot}=[10,30]\kpc$, and analyse only those sources with Galactic latitudes $|b|>20\deg$. We note that LMS-1 was originally detected in our regular \texttt{STREAMFINDER} run on \Gaia EDR3 (using an identical procedure described in \citetalias{Ibata_GaiaEDR3}). However, after this, we conducted a tailored run to better map out this stream (using the parameters described above)\footnote{For instance, the prime reason for adopting the [Fe/H]=$-2.2$ model was that when we cross-matched the initially found LMS-1 members with the spectroscopic surveys, we inferred a metallicity of $\sim -2.2$~dex.}.

Figure~\ref{fig:Fig_SF_map} shows the all-sky map of \Gaia EDR3 stars that exhibit stream-like behavior with significance $>10\sigma$ (as per the \texttt{STREAMFINDER} analysis). The sample shown corresponds to the selection in proper motion as $-3.0\masyr <\mu^{*}_{\ell} < -0.3\masyr$ and $-4.0\masyr <\mu_{b} < 1.0\masyr$, as this selection was appropriate to highlight the structures of interest. This map shows several structures, including some of the previously known streams such as ``Orphan'' \citep{Koposov2019}, ``Galactic Anticenter stellar structure(s)'' \citep{Newberg_GASS_2002, Ibata_GASS_2003, Slater2014} and parts of Sagittarius \citep{Ramos2020, Ibata_Sgr2020}. The structure at $(\ell,b)\sim(-80\deg,-70\deg)$ is possibly the southern part of the Cetus stream, but it is hard to conclude on its true nature at this stage. Among these objects, the LMS-1 stream stands out as one of the most striking structures. LMS-1 lies towards the north of the Galactic bulge between $\ell=[-50\deg,50\deg],b=[40\deg, 80\deg]$, and is located at a heliocentric distance of $d_{\odot}\sim 20\kpc$ (as inferred from the uncertainty-weighted average mean parallax), and at a galactocentric distance of $\sim 15\kpc$. The heliocentric distance and the angular dispersion of the stream imply that LMS-1 has a physical width of (Gaussian) dispersion of $\sim700\pc$. 

\begin{figure}
\begin{center}
\vspace{-0.3cm}
\includegraphics[width=\hsize]{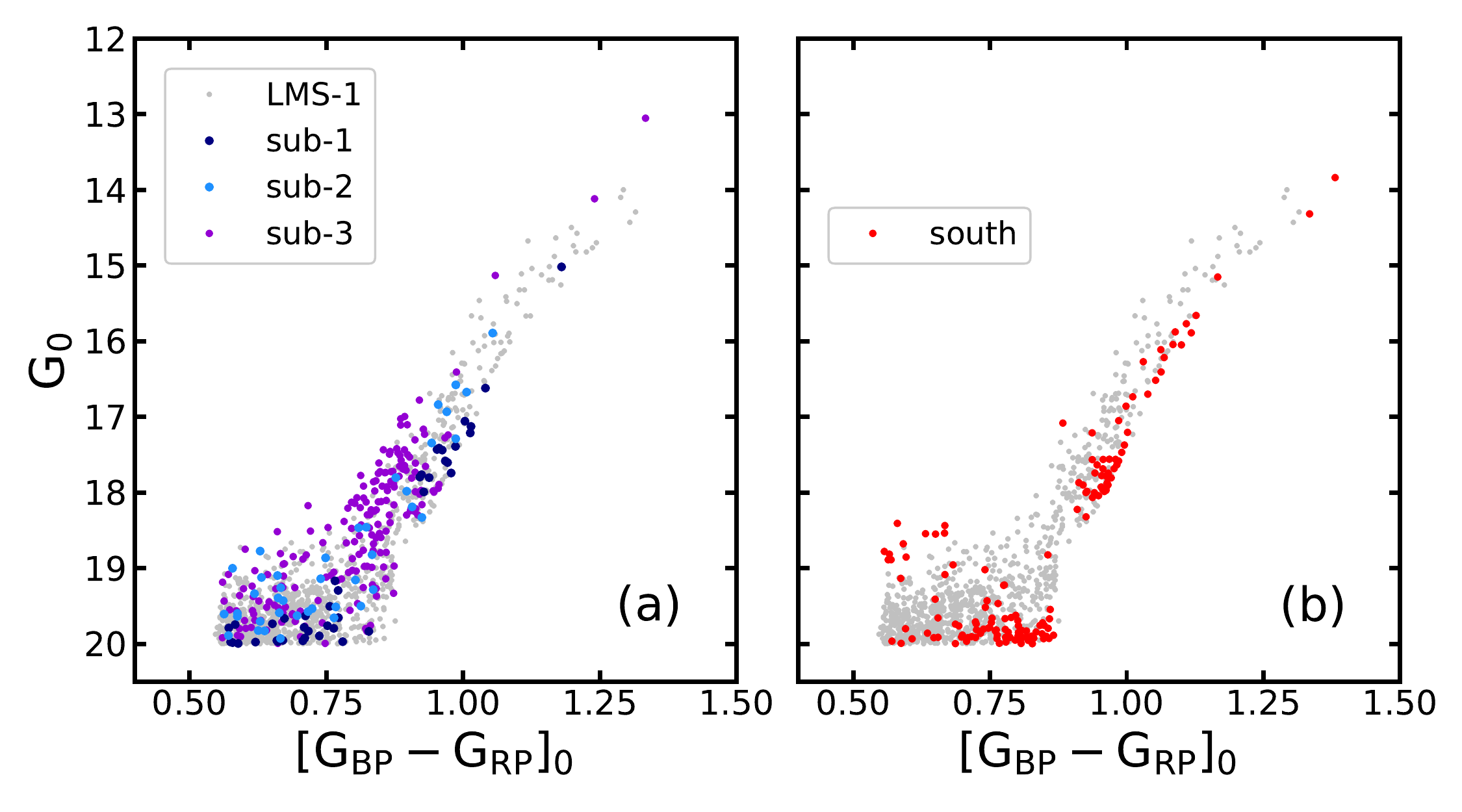}
\end{center}
\vspace{-0.5cm}
\caption{Comparing the color-magnitude distribution of LMS-1 with sub-1, sub-2, sub-3 and the southern sub-structure visible in Figure~\ref{fig:Fig_SF_map}. The data are corrected for dust extinction in \Gaia's $G$ vs $G_{\rm BP}-G_{\rm RP}$ photometry.}
\label{fig:Fig_CMD_LMS1_comps}
\end{figure}

In Figure~\ref{fig:Fig_SF_map}, we also observe some other sub-structures that are likely associated with LMS-1. For instance, we note three newly-found off-stream features that are located adjacent to LMS-1 on sky, labelled as sub-1, sub-2 and sub-3. Using the uncertainty-weighted average mean parallax, we infer the distance of sub-2 and sub-3 as $d_{\odot}\sim21\kpc$ and $8\kpc$, respectively. For sub-3, the \texttt{STREAMFINDER} finds a distance solution of $\sim16\kpc$. These two very different distance estimates for sub-3 may be due an intrinsically large distance range along the line of sight, but it may also simply reflect the difficulty of measuring distances for such low surface brightness structures. Sub-1 contains very few stars to provide any useful distance estimate. Color magnitude distributions (CMDs) of these features are compared with LMS-1 in Figure~\ref{fig:Fig_CMD_LMS1_comps}, and they show some similarities in their stellar population. Moreover, we also detect an agglomeration of stars in the southern Galactic sky around  $(\ell,b) \sim (-40\deg,-60\deg)$. The uncertainty-weighted average mean parallax of these stars provides a distance estimate of $d_{\odot}\sim19\kpc$, similar to that of LMS-1. The CMD of this southern sub-structure is compared with LMS-1 in Figure~\ref{fig:Fig_CMD_LMS1_comps}b (the sub-giant branch of this candidate feature seems cut-off by the limiting magnitude at $G_0=20$). The fact that many of these sub-structures have similar CMDs and distances as that of LMS-1, and that they also possess similar on-sky curvature as LMS-1, may be indicating that they originated from the same parent galaxy. To examine this scenario further, we examine the kinematic information of these sub-structures in Section~\ref{sec:Nbody_model}.   

The detected LMS-1 stream does not necessarily represent the entire underlying structure. This is because just like any other stream finding algoirthm, even \texttt{STREAMFINDER} does not guarantee the recovery of all the relics of the progenitor (e.g., all the wraps of the accreted dwarf and all the associated components). Assessing this completeness in the LMS-1 detection by \texttt{STREAMFINDER} is beyond the scope of this paper. Secondly, in order to investigate the possible bias due to the choice of the stellar population model, we re-processed the \Gaia data with \texttt{STREAMFINDER} using a slightly metal rich template of [Fe/H]$=-1.5$ (rest of the parameters were kept fixed to their original values). We infer that our final results do not vary with this change, and conclude that the metallicity selection-bias is not very significant.

\begin{figure}
\begin{center}
\vspace{-0.3cm}
\includegraphics[width=0.95\hsize]{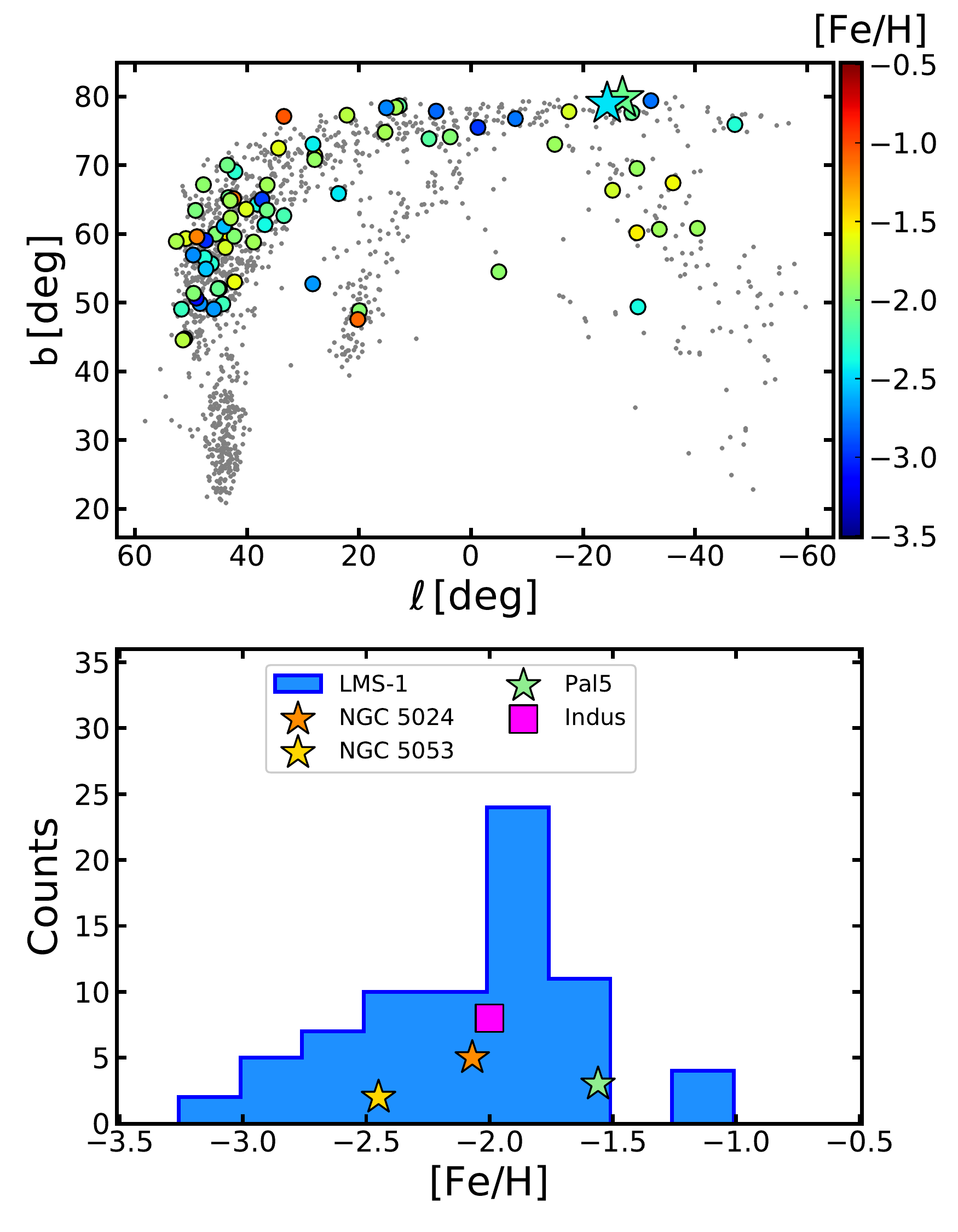}
\end{center}
\vspace{-0.5cm}
\caption{The metallicity distribution of the LMS-1 stream. The top panel focuses around the northern region of Figure~\ref{fig:Fig_SF_map}, where the \Gaia\ stars are shown in gray. Among these stars, those possessing spectroscopic [Fe/H] measurements are color coded as per their [Fe/H] values. The globular clusters NGC~5053 and NGC~5024 are also shown (with ``star'' markers). The bottom panel shows the 1D distribution corresponding to all of these [Fe/H] measurements. The panel also shows the [Fe/H] values of those objects that are dynamically associated with LMS-1 (see text).}
\label{fig:Fig_FeH}
\end{figure}

\subsection{Metallicity of LMS-1}\label{subsec:FeH_LMS1}

To measure the metallicity ([Fe/H]) of LMS-1 stars, we made the selection of the northern part of the structure as shown in the top panel of Figure~\ref{fig:Fig_FeH} and cross-matched these sources with public spectroscopic surveys. We found $N=41$ matches in LAMOST DR6 \citep{Zhao2012}, $N=18$ in SDSS/SEGUE DR10 \citep{Yanny2009} and $N=2$ in APOGEE DR14 \citep{Majewski_APOGEE_2017}. Furthermore, we discovered that there are a significant number of spectra for LMS-1 stars in LAMOST DR6 that do not have stellar parameters in their catalogue. Additionally, the DR6 parameters saturate at [Fe/H]$=-2.5$ due to limitations in their pipeline. We therefore re-analysed all LMS-1 LAMOST spectra to gain additional member stars and to derive more accurate metallicities. The spectra were analysed in the spectral range of $3800-5500$~\AA~using the full-spectrum fitting package ULySS \citep{Koleva09}, with the latest version of the MILES library model interpolator \citep{prugniel11, sharma16}. The effective temperature, surface gravity, metallicity and radial velocity are determined simultaneously in the fit. For giant stars, metallicities can be derived reliably down to [Fe/H]$=-3.0$. We tested our method on the very metal-poor LAMOST sample by \cite{Li_LAMOSTDR32018} and find metallicities consistent with theirs (see Appendix~\ref{appendix:FeH_comparison}). We derive stellar parameters for $N=57$ stars, of which $N=22$ do not have parameters in LAMOST DR6. Most of the missing stars have [Fe/H]$<-2.5$. The internal fitting uncertainties on [Fe/H] are $0.1-0.25$~dex and $5-15\kms$ on the radial velocities, depending on the signal-to-noise ratio. The $rms$ between our metallicities and radial velocities and those from the LAMOST DR6 catalogue is $0.15$~dex and $7\kms$, respectively. Assuming both determinations contribute equally to the $rms$, we add $rms/\sqrt{2}$ in quadrature to our uncertainties to include an estimate of the external uncertainties. Henceforth, all the LAMOST parameters refer to our derived values. 

As shown in Figure~\ref{fig:Fig_FeH}, LMS-1 is a very metal poor stream with mean [Fe/H]$=-2.09\pm0.05$, and a (Gaussian) intrinsic dispersion of $\sigma_{\rm [Fe/H]}=0.42\pm0.04$. These values were computed following our own Metropolis–Hastings based MCMC algorithm, where the log-likelihood function is taken to be:
\begin{equation}
\ln \mathcal{L} = \sum_{\rm data} \Big[ -\ln \sigma_{\rm [Fe/H]_{obs}} -0.5 \Big(\dfrac{\rm [Fe/H]_0-[Fe/H]_{\rm d}}{\sigma_{\rm [Fe/H]_{obs}}}\Big)^2 \Big]\,.
\end{equation}
Here, ${\rm [Fe/H]_{d}}$ is the measured metallicity, and the Gaussian dispersion $\sigma_{\rm [Fe/H]_{obs}}$ is the sum in quadrature of the intrinsic dispersion of the model together with the observational uncertainty of each data point ($\sigma^2_{\rm [Fe/H]_{obs}}=\sigma^2_{\rm [Fe/H]} + \sigma^2_{\rm [Fe/H]_d}$). For this inference, we impose a very conservative metallicity cut, selecting only those stars with [Fe/H]$<-1$ so as to retain a maximal population of LMS-1. Moreover, we note that the median of this [Fe/H] distribution is $-1.96$~dex. Our [Fe/H] measurement of LMS-1 is consistent with the value reported in \citet{Yuan2020}. A non-zero and large value of $\sigma_{\rm [Fe/H]}$ implies that LMS-1 is produced from a dwarf galaxy. For the stars in sub-1, sub-2 and sub-3, we found either zero or very few cross-matches, and therefore it is difficult to infer whether their [Fe/H] is similar to that of LMS-1.

Figure~\ref{fig:Fig_FeH} also shows the [Fe/H] measurements of the globular clusters NGC~5053 \citep{Boberg_NGC5053_2015}, NGC~5024 \citep{Boberg_NGC5024_2016} and Pal~5 \citep{Koch2017}, and the ``Indus'' stellar stream \citep{Shipp2018, Ji_S5_2020}. Note that all of these objects fall within the [Fe/H] range of LMS-1. An additional reason for highlighting these particular objects is that they are also strongly linked with LMS-1 in the dynamical action-energy space (as we show in Section~\ref{sec:ActionEnergy}). 

Next, we investigate the alpha abundances for LMS-1 stars. The one star in APOGEE has $[\rm{\alpha/M}]=0.38\pm0.06$ and [Mg/Fe]$=0.31\pm0.07$ (this star has [Fe/H]=$-2.09\pm0.09$). We also cross-matched the above dataset with the \cite{Xiang_LAMOSTDR5_2019} catalogue of LAMOST DR5. With $N=51$ positive cross-matches we infer $\rm{[\alpha/Fe]}\sim 0.25$ (similar to the one obtained by \citealt{Naidu2020} for the Wukong structure). However, we are skeptical about this value because (1) \cite{Xiang_LAMOSTDR5_2019} analysis of LAMOST DR5 has not been carefuly tested for very metal-poor stars, especially for the stars with [Fe/H]$<-1.5$, and (2) we found that for very metal-poor stars from \citet{Li_LAMOSTDR32018}, the [Fe/H] values in the \cite{Xiang_LAMOSTDR5_2019} catalogue are $\sim0.4$~dex higher than the literature values (see Appendix~\ref{appendix:FeH_comparison}). Therefore, we conclude that the alpha abundances (and even metallicities) from the \cite{Xiang_LAMOSTDR5_2019} catalogue should likely not be used for very metal-poor stars, such as those in LMS-1.

It is important to note that our estimate of ${\rm \langle [Fe/H]\rangle}$ for LMS-1, and also the median, is $\sim0.4$~dex lower than the median of [Fe/H] distribution reported for Wukong by \cite{Naidu2020}. This is strange given that these two objects likely represent the same structure, since they possess very similar dynamical properties (we show this for LMS-1 below). A possible reason for this [Fe/H] discrepancy could be systematic differences in the metallicity scales of our analysis and that of the H3 survey (which \citealt{Naidu2020} used); we briefly discuss this in Appendix~\ref{appendix:FeH_comparison}. Another reason could be the difference in the procedures to select the LMS-1 member stars. The spectroscopic sample selected from \texttt{STREAMFINDER} is purely based on the information of phase space and stellar population. However, the Wukong sample is first selected from a box in the ($L_z, E$) space, and a further cut is made based on the break in [Fe/H] distribution function, which possibly have more contamination by (more metal-rich) halo stars, leading to a higher average metallicity.

\begin{figure*}
\begin{center}
\vspace{-0.3cm}
\includegraphics[width=\hsize]{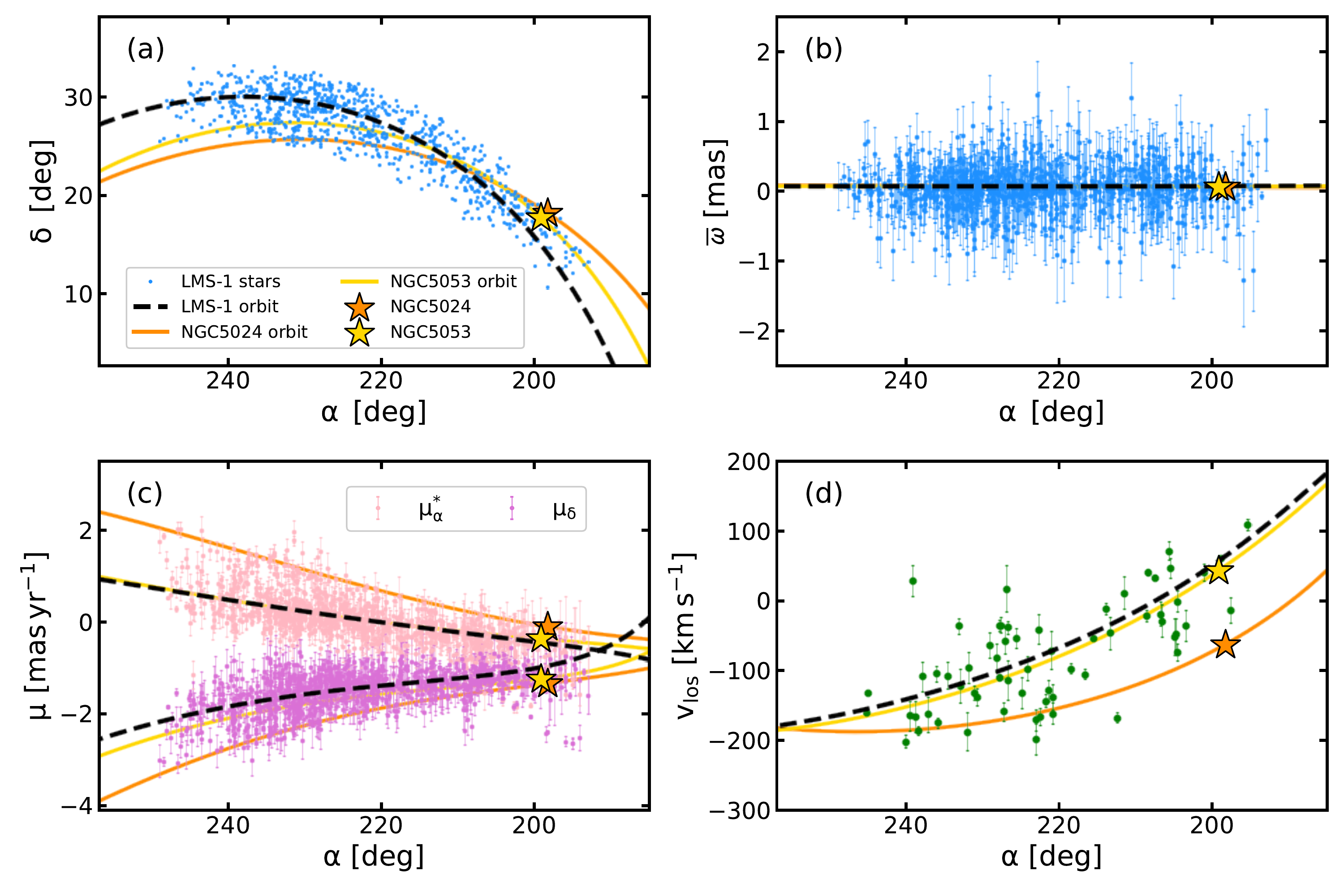}
\end{center}
\vspace{-0.5cm}
\caption{Orbit-fit to the LMS-1 stream. From (a) to (d) the panels compare the best orbit-fit model to the data in position, parallax, proper motion and line-of-sight velocity space. These panels also show the phase-space positions and orbits of the globular clusters NGC~5024 and NGC~5053.} 
\label{fig:Fig_Orbit_fit}
\end{figure*}
\section{Orbit of LMS-1}\label{sec:orbit}

The orbit of a stellar stream informs about the path along which the parent galaxy was accreted, got tidally stripped, and thereby contributed to populating the stellar halo. To constrain the orbit of LMS-1, we fit orbits to the stars shown in Figure~\ref{fig:Fig_Orbit_fit}. These sources correspond to a $10\deg$ wide selection made around the stream path in the northern galactic sky, that also allows us to isolate LMS-1 stars of interest from the neighbouring sub-structures. This selection results in a sample of $N=877$ stars, and all of these stars possess the 5D astrometric measurements of positions, parallaxes and proper motions. To obtain the line-of-sight (los) velocities $(v_{\rm los})$ we resort to the aforementioned cross-matches with spectroscopic surveys. We find that only $N=63$ of these stars possess $v_{\rm los}$ measurements. To deal with the missing $v_{\rm los}$ information for the remaining stars, we set them all to $v_{\rm los}= 0\kms$ with a $10^4\kms$ Gaussian  uncertainty\footnote{The results are almost identical if instead an uncertainty of $10^3\kms$ is assumed. The choice of adopting such a large uncertainty is effectively imposing a prior that the stars must be located in the local universe. See \cite{Malhan_GalPot_2019}.}. The LMS-1 stars that possess spectroscopic $v_{\rm los}$ measurements are listed in Table~\ref{tab:table1}.

To compute orbits, we adopted the Galactic potential model of \cite{McMillan2017}. This is an axisymmetric model comprising a bulge, disk components and an NFW halo. To set this Galactic potential model, and to compute orbits, we make use of the \texttt{galpy} module \citep{Bovy2015}. Moreover, to transform the orbits from the Galactocentric frame into the Heliocentric frame (required to implement equation~\ref{eq:Loglikehood_Potential1}) we adopt the same values of the Sun's Galactocentric distance and the Sun's galactic velocity as described in Section~\ref{sec:Data}. Finally, to obtain the best orbit model for LMS-1, we survey the parameter space using our own Metropolis-Hastings based MCMC algorithm, where the log-likelihood of each data point $i$ is defined as
\begin{equation}\label{eq:Loglikehood_Potential1}
\ln \mathcal{L}_i = -\ln ((2\pi)^{5/2}\sigma_{\rm sky} \sigma_{\overline{\omega}} \sigma_{\mu_{\alpha}} \sigma_{\mu_{\delta}} \sigma_{v_{\rm los}}) +\ln N -\ln D,
\end{equation}
where 
\begin{equation}
\begin{aligned}
N &= \prod_{j=1}^5 (1-e^{-R^2_j/2}) \, , \\
D &= \prod_{j=1}^5 R^2_j \, , \\
R_1^2 &= \dfrac{\theta^2_{\rm sky}}{\sigma^2_{\rm sky}} \, , \\
R_2^2 &= \dfrac{(\overline{\omega}^{\rm d} - \overline{\omega}^{\rm o})^2}{\sigma^2_{\overline{\omega}}} \, , \\
R_3^2 &= \dfrac{(\mu^{\rm d}_{\rm \alpha} - \mu^{\rm o}_{\rm \alpha})^2}{\sigma^2_{\mu_{\alpha}}} \, , \\
R_4^2 &= \dfrac{(\mu^{\rm d}_{\rm \delta} - \mu^{\rm o}_{\rm \delta})^2}{\sigma^2_{\mu_{\delta}}} \, , \\
R_5^2 &= \dfrac{(v^{\rm d}_{\rm los} - v^{\rm o}_{\rm los})^2}{\sigma^2_{v_{\rm los}}} \, .\\
 \\
\end{aligned}
\end{equation}

Here, $\theta_{\rm sky}$ is the on-sky angular difference between the orbit and the data point, $\overline{\omega}^{\rm d}, \mu^{\rm d}_{\rm \alpha}, \mu^{\rm d}_{\rm  \delta}$  and $v^{\rm d}_{\rm los}$ are the measured data parallax, proper motion and los velocity, with the corresponding orbital model values marked with ``$o$''. The Gaussian dispersions $\sigma_{\rm sky}, \sigma_{\overline{\omega}}, \sigma_{\mu_{\alpha}}, \sigma_{\mu_{\delta}}, \sigma_{v_{\rm los}}$ are the sum in quadrature of the intrinsic dispersion of the model and the observational uncertainty of each data point. The reason for avoiding the standard log-likelihood function, and adopting this ``conservative formulation'' of \cite{sivia1996data}, is to lower the contribution from outliers that could be contaminating our data. The success of this modified log-likelihood equation in fitting streams has been demonstrated in \cite{Malhan_GalPot_2019}.

\begin{figure}
\begin{center}
\vspace{-0.3cm}
\includegraphics[width=\hsize]{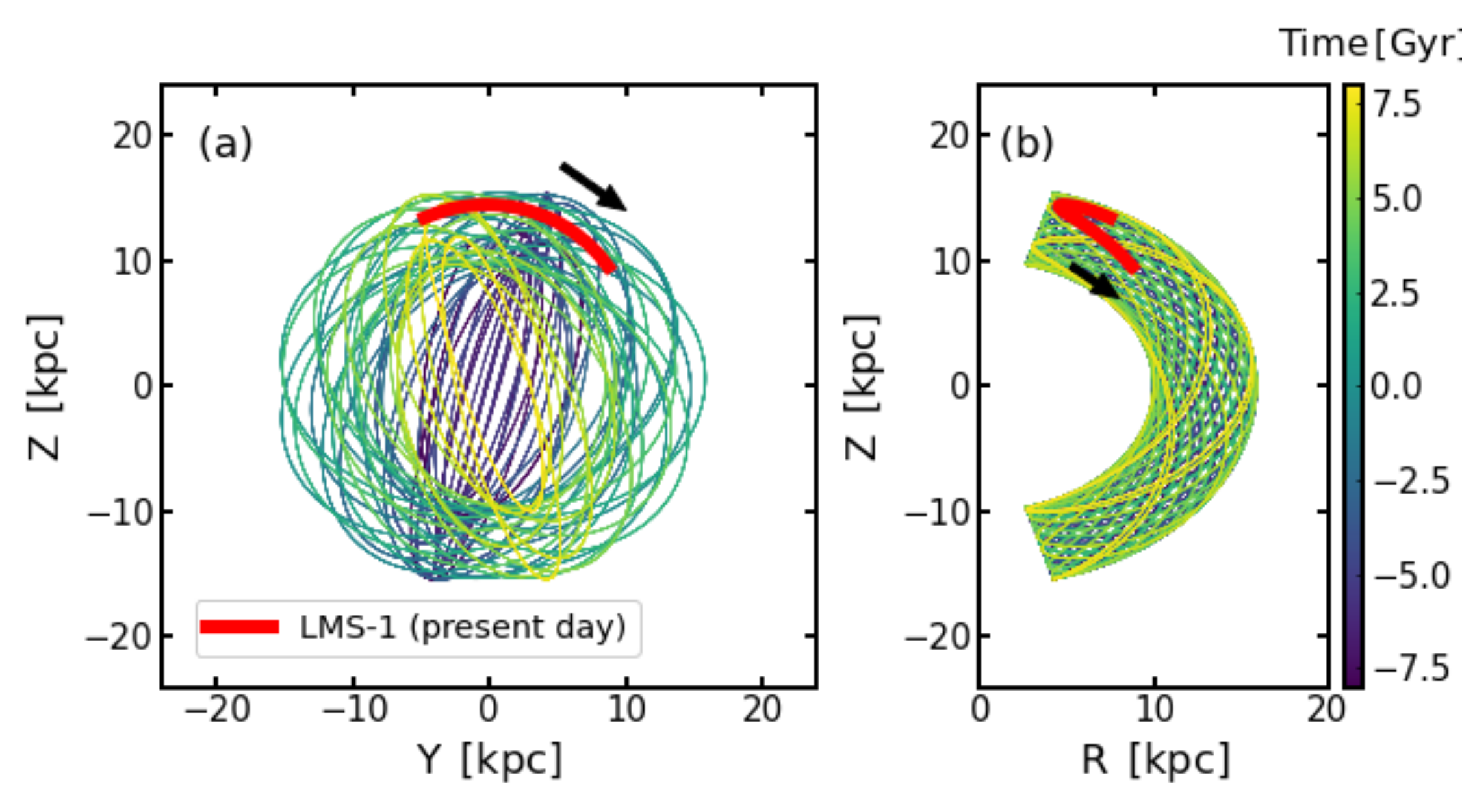}
\end{center}
\vspace{-0.5cm}
\caption{The orbit of the LMS-1 stream in Galactocentric coordinates. (a) The orbit is presented in the Galactic y−z plane, and is color coded by integration time (where Time=0 represents the present day). For perspective the present day location of LMS-1 is shown in red. (b) Same as (a) but in the Galactic R−z plane. The arrows represent the direction of motion of LMS-1.} 
\label{fig:Fig_Orbit_Long}
\end{figure}

The best-fit orbit for LMS-1 is shown in Figure~\ref{fig:Fig_Orbit_fit}. We find this orbit to be very circular (eccentricity$\sim0.20$), slightly prograde (z-component of angular momentum is $L_z\sim -700\kms \kpc$) and quite polar (with an inclination of $\sim75\deg$ to the Galactic plane). Note that our inferred value of LMS-1's inclination is slightly lower than the previous estimate of $80\deg$ by \cite{Yuan2020}. This new estimate of the LMS-1's inclination is similar to that of the Sagittarius stream ($\sim76\deg$, \citealt{Yuan2020}), and significantly higher than that of the Cetus stream ($60\deg$, \citealt{Chang_2020_Cetus}). Furthermore, LMS-1's orbit has a small apocentre of $r_{\rm apo}\sim16\kpc$, a pericentre of $r_{\rm peri}\sim10\kpc$ and a maximum height above the Galactic plane of $z_{\rm max}\sim 16\kpc$. Some of these orbital parameters were previously computed by \cite{Yuan2020} and here we find slightly different values. Furthermore, the best-fit orbit also indicates that the heliocentric distance of LMS-1 is $d_{\odot}\sim 16\kpc$ ($\sim20\%$ smaller from the one we obtained above using parallaxes). The orbit of LMS-1 is also shown in Figure~\ref{fig:Fig_Orbit_Long} in the Galactocentric Cartesian system.

Figure~\ref{fig:Fig_Orbit_fit} also shows the remarkable coincidence of the LMS-1's orbit with the positions, proper motions and velocities of the globular clusters NGC~5053 and NGC~5024 (values for the globular cluster taken from \citealt{Baumgardt2019}). This coincidence was also previously noted by \cite{Yuan2020, Naidu2020}. Moreover, the heliocentric distance of these two clusters ($\sim17-18\kpc$) is also very similar to that of LMS-1. To facilitate visual comparison, we also show the orbits of these clusters in Figure~\ref{fig:Fig_Orbit_fit}. Admittedly, the coincidence of NGC~5024 with LMS-1 is not so obvious in $v_{\rm los}$, as one can note a large offset of $\sim 100\kms$ between their orbits. However, given the large velocity spread of LMS-1 that is apparent in Figure~\ref{fig:Fig_Orbit_fit}, it is possible that these two systems are indeed dynamically associated. This phase-space overlap strongly indicates that NGC~5053 and NGC~5024 were possibly accreted within the LMS-1's parent galaxy, and now all of these objects are orbiting in the Milky Way along very similar trajectories. We explore this scenario further in Section~\ref{sec:ActionEnergy}. 

\begin{table}
\centering
\caption{Stars of LMS-1 with spectroscopic observations (details of only first $10$ stars are shown).}
\label{tab:table1}
\begin{tabular}{|l|l|l|l|l|}

\hline
\hline
{\bf Survey} &  {\bf R.A}. & {\bf Dec.} & {\bf $v_{\rm los}$} & {\bf [Fe/H]} \\
&[deg] & [deg] & [$\kms$]& [dex]\\
\hline
\hline

2  &  195.3385  &  13.22992  &  109.51 $\pm$ 8.03  &  -2.32 $\pm$ 0.13  \\
2  &  197.53424  &  17.42584  &  -14.22 $\pm$ 17.59  &  -2.79 $\pm$ 0.17  \\
1  &  198.938  &  16.12697  &  59.64 $\pm$ 0.11  &  -2.09 $\pm$ 0.09  \\
2  &  200.9852  &  17.4919  &  40.47 $\pm$ 11.07  &  -1.68 $\pm$ 0.13  \\
2  &  203.35826  &  18.0979  &  -35.01 $\pm$ 21.85  &  -2.78 $\pm$ 0.2  \\
2  &  204.44497  &  22.74855  &  -73.71 $\pm$ 11.49  &  -2.12 $\pm$ 0.16  \\
2  &  204.48571  &  21.20221  &  -1.6 $\pm$ 15.05  &  -2.82 $\pm$ 0.16  \\
2  &  204.71231  &  22.77687  &  -48.22 $\pm$ 21.85  &  -1.85 $\pm$ 0.27  \\
2  &  204.93202  &  23.05065  &  -51.28 $\pm$ 25.93  &  -2.74 $\pm$ 0.2  \\
2  &  205.48435  &  18.43294  &  46.62 $\pm$ 14.54  &  -2.97 $\pm$ 0.15  \\
3  &  205.72079  &  20.09672  &  71.51 $\pm$ 12.22  & -  \\

\hline
\hline
\end{tabular}
\tablecomments{The first column provides the names of the source survey (APOGEE=1, LAMOST=2, SDSS/SEGUE=3). Columns $2-3$ list the equitorial coordinates R.A. and Declination, respectively. Columns $4-5$ provide the measured line-of-sight velocities and measured metallicities.}
\end{table}

\vspace{1.0cm}
\section{Velocity dispersion of LMS-1}\label{sec:vel_disp}

The velocity dispersion of stellar streams informs us on the nature of their progenitors (for instance, whether the system in question was of high or low mass), and it also serves as an observational constraint to construct dynamical models. Moreover, the estimate of a stream's velocity dispersion provides an approximate extent to which the stream stars could be spread out in phase-space. 

We estimate the velocity dispersion of LMS-1 using the best-fit orbit model obtained in Section~\ref{sec:orbit}. Our method is similar to that adopted in \cite{Malhan_GalPot_2019}. We decompose the internal (3-dimensional) velocity dispersion ($\sigma_{\rm v, int}$) as the sum
\begin{equation}
\sigma^2_{\rm v, int}=\sigma^2_{\rm v_T,int}+\sigma^2_{\rm vlos,int}\,,
\end{equation}
\noindent where $\sigma_{\rm v_T,int}$ and $\sigma_{\rm vlos,int}$ are the tangential and the line-of-sight components of the velocity dispersion, respectively. Therefore, to compute $\sigma_{\rm v, int}$, we first independently estimate these two components of velocity dispersion. 

To measure $\sigma_{\rm v_T,int}$, we take advantage of \Gaia's precise proper motion measurements and assume that $\sigma^2_{\rm v_T,int} = 2\sigma^2_{\rm v_\alpha,int} = 2\sigma^2_{\rm v_\delta,int}$. Here, the $\alpha$ and $\delta$ correspond to components along the directions of Right Ascension and Declination, respectively. With this, the  log-likelihood function to compute $\sigma_{\rm v_T,int}$ is taken to be:
\begin{equation}\label{eq:Vel_disp_no_contam}
\ln \mathcal{L}_1 = \sum_{\rm data} \Big[  -\ln(2\pi \sigma^{\rm obs}_{\rm v_{\alpha}}\sigma^{\rm obs}_{\rm v_{\delta}}\sqrt{1-\rho^2}) +\ln N -\ln D \Big]\,,
\end{equation}
\noindent where
\begin{equation}
\begin{aligned}
N &= \prod_{j=1}^3 (1-e^{-\frac{R^2_j}{2(1-\rho^2)}}) \, , \\
D &= \prod_{j=1}^3 R^2_j \, , \\
R_1^2 &= \Big(\dfrac{v^{\rm o}_{\alpha}-v^{\rm d}_{\alpha}}{\sigma^{\rm obs}_{\rm v_{\alpha}}}\Big)^2\,,\\
R_2^2 &= \Big(\dfrac{v^{\rm o}_{\delta}-v^{\rm d}_{\delta}}{\sigma^{\rm obs}_{\rm v_{\delta}}}\Big)^2\,,\\
R_3^2 &= -2\rho \Big(\dfrac{v^{\rm o}_{\alpha}-v^{\rm d}_{\alpha}}{\sigma^{\rm obs}_{\rm v_{\alpha}}}\Big)\Big(\dfrac{v^{\rm o}_{\delta}-v^{\rm d}_{\delta}}{\sigma^{\rm obs}_{\rm v_{\delta}}}\Big)  \,.\\
\end{aligned}
\end{equation}
Here, $v^{\rm d}_{\alpha},v^{\rm d}_{\delta}$ are the observed tangential velocity components (calculated by multiplying the proper motion measurement with the orbit model distance), and the corresponding orbital model values are marked with ``o''. The Gaussian dispersions $\sigma_{\rm v_{\alpha, obs}},\sigma_{\rm v_{\delta, obs}}$ are the sum in quadrature of the observational uncertainty of each data point together with the intrinsic dispersion (i.e., $\sigma^2_{\rm obs} = \sigma^2_{\rm d} + \sigma^2_{\rm int}$, and the quantity of interest is $\sigma_{\rm int}$). The parameter $\rho$ takes into account the measured correlation in proper motions $C(=$\texttt{pmra\_pmdec\_corr} in the \Gaia dataset) such that 
\begin{equation}\label{eq:rho}
\rho = \frac{C\sigma^{\rm d}_{\mu_\alpha}\sigma^{\rm d}_{\mu_\delta}}{\sqrt{(\sigma^{\rm d}_{\mu_\alpha})^2+(\sigma^{\rm int}_{\mu_\alpha})^2}  \sqrt{(\sigma^{\rm d}_{\mu_\delta})^2  +(\sigma^{\rm int}_{\mu_\delta})^2}} \, .
\end{equation}
To estimate $\sigma_{\rm vlos, int}$, we use the spectroscopic $v_{\rm los}$ measurements, and the log-likelihood function is taken to be:
\begin{equation}
\ln \mathcal{L}_2 = \sum_{\rm data} \Big[  -\ln(\sqrt{2\pi} \sigma_{\rm v_{los,obs}}) + \ln N -\ln D \Big]\,,
\end{equation}
\noindent where
\begin{equation}
\begin{aligned}
D &=\Big(\dfrac{v^{\rm o}_{\rm los}-v^{\rm d}_{\rm los}}{\sigma_{\rm v_{los,obs}}}\Big)^2 \, , \\
N &= (1-e^{-\frac{D}{2}})\,.
\end{aligned}
\end{equation}
Here, $v^{\rm d}_{\rm los}$ is the observed los velocity and the corresponding orbital model values are marked with ``o''. The Gaussian dispersion $\sigma_{\rm v_{los, obs}}$ are the sum in quadrature of the intrinsic dispersion of the model together with the observational uncertainty of each data point ($\sigma^2_{\rm obs}=\sigma^2_{\rm int} + \sigma^2_{\rm d}$). 

\begin{figure}
\begin{center}
\vspace{-0.3cm}
\includegraphics[width=\hsize]{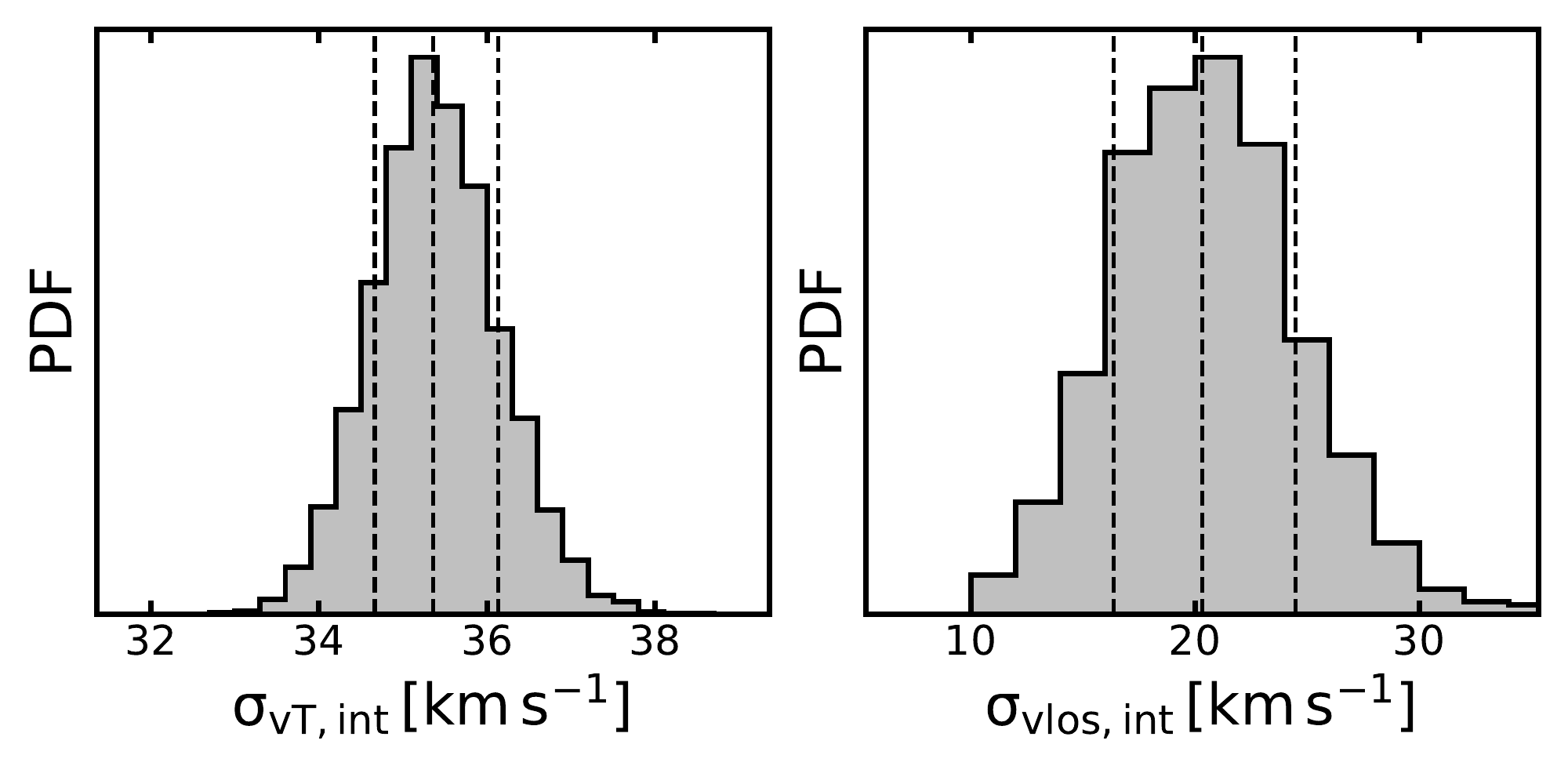}
\end{center}
\vspace{-0.5cm}
\caption{Velocity dispersion of the LMS-1 stream along the tangential (left) and line-of-sight (right) directions. The dashed lines correspond to quantiles (0.16,0.50,0.84).}
\label{fig:Fig_vel_disp}
\end{figure}

The resulting velocity dispersion distribution is shown in Figure~\ref{fig:Fig_vel_disp}. We infer $\sigma_{\rm v_T, int} =35.4^{+ 0.8 }_{- 0.7 }\kms$ and $\sigma_{\rm vlos, int} = 20\pm 4\kms$. These values together imply that the total 3-dimensional velocity dispersion of LMS-1 is $\sigma_{\rm v,int}=40.9\pm 2.1\kms$. Such a large velocity dispersion has not been previously measured for any stellar stream in the Milky Way. Even the massive Sagittarius stream has a line of sight velocity dispersion of only $\sim 13\kms$ \citep{Gibbons2017}, which is slightly smaller than $\sigma_{\rm vlos, int}$ of LMS-1. It is important to note that if the detected LMS-1 is dynamically colder than the actual underlying substructure, then this velocity dispersion value only provides a lower limit. We discuss the possible sources of such a large velocity dispersion of LMS-1 in Section~\ref{sec:Discussion_Conclusion}.

\begin{figure*}
\begin{center}
\includegraphics[width=0.8\hsize]{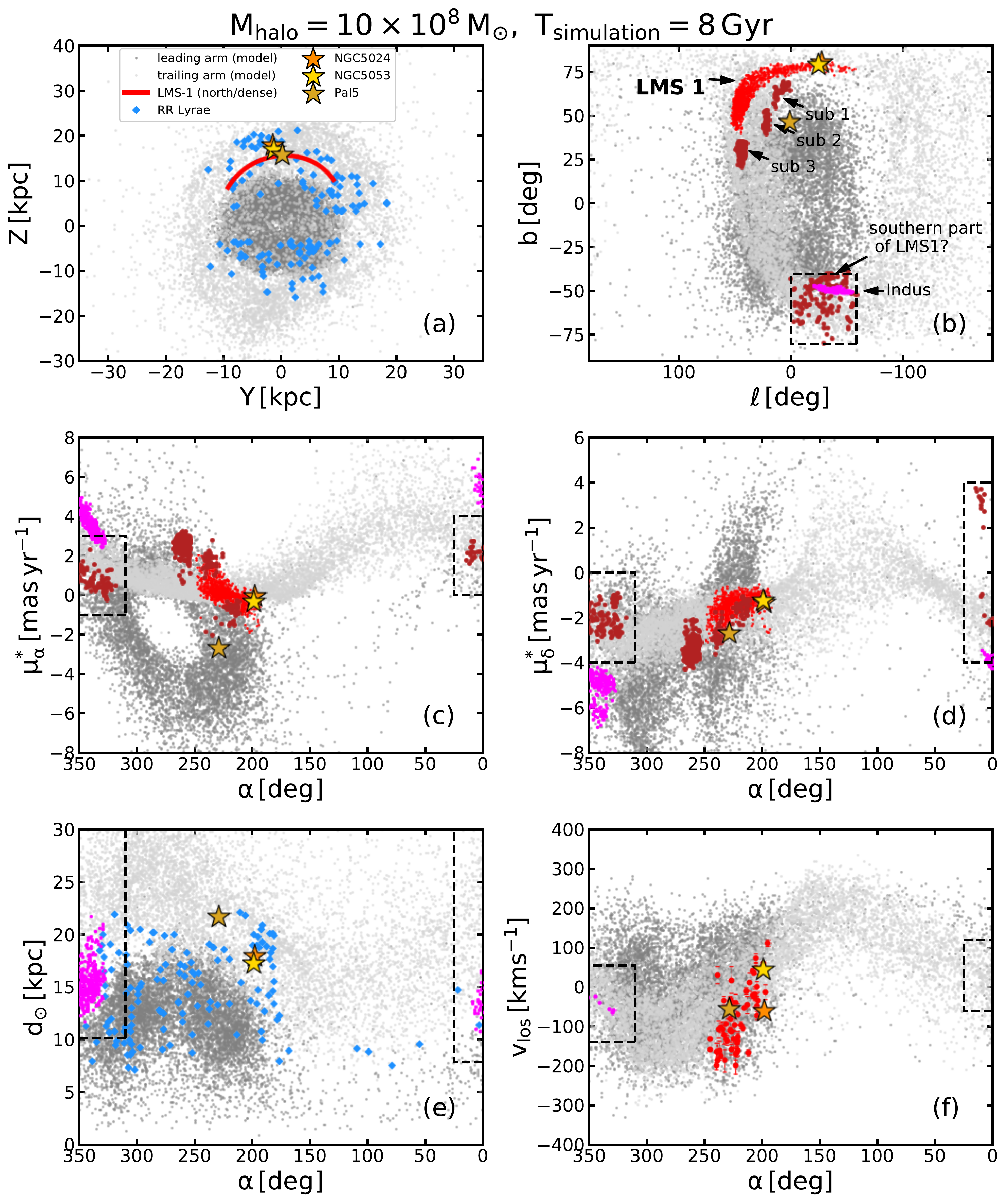}
\end{center}
\vspace{-0.5cm}
\caption{Comparing phase-space distribution of simulation particles with the data of LMS-1, other sub-structures, globular clusters NGC~5024, NGC~5053 and Pal~5, and the Indus stream. Panel (a) and (b) show Galactocentric Y-Z coordinate and Galactic sky coordinates, respectively. From (c) to (f) the panels show proper motion along Right Ascension, proper motion along Declination, heliocentric distance and line-of-sight velocity as functions of Right Ascension. Panels (a) and (e) also show the RR Lyrae stars found along the LMS-1's orbit. The simulation is also able to reproduce the southern agglomeration of stars, at least in the position and the proper motion space (the rectangle highlights the simulated particles that overlap with the southern stars).}
\label{fig:Fig_compare_phase-space}
\end{figure*}
\section{Tracing LMS-1 in Gaia's RR Lyrae catalogue}\label{sec:RRLyare_sel}
 
RR Lyrae provide a means to determine very accurate distances of stellar structures, which can otherwise be hard to estimate even from \Gaia\ parallaxes (at least for distant halo objects). Especially for streams, finding their associated RR Lyrae can provide distance anchors to the stream in the 3D spatial volume of the halo, and this assists in improving dynamical models. Here we use the best-fit orbit of LMS-1 as the reference to create a catalog of LMS-1 RR Lyrae stars.

The RR Lyrae data that we use is the same described in \cite{Ibata_Sgr_2020}, that is based on the sources identified in the \texttt{gaiadr2.vari\_rrlyrae} catalog of \Gaia DR2 \citep{Clementini2019}. To select the LMS-1 RR Lyrae candidates from this dataset, we use the best-fit orbit model integrated from Time=$-2$ to $+2$ Gyr (with Time=0 representing the present day location of LMS-1). We use the orbit's positional and proper motion information, and perform sigma-clipping to select only those RR Lyrae stars that lie within $2\sigma$ of the model (we do not use the orbit's distance or line-of-sight velocity information). For this selection, we also use the inferred values of LMS-1's physical width and its tangential velocity dispersion. Moreover, we select only those stars that lie more than $3\kpc$ above or below the Galactic plane (to avoid the contamination from the stellar disk), range in distance from $d_{\odot}=7-22\kpc$ (comparable to the range between perigalacticon and apogalacticon of LMS-1), and for which [Fe/H]$<-1$ (if metallicity is available). Furthermore, we also avoid all those RR Lyrae sources that lie in the regions containing NGC~5024, NGC~5053, Pal~5, NGC~5272, IC 4499, LMC, SMC and also the Sagittarius stream.

Following these selection criteria, we obtain a sample of $N=134$ RR Lyrae stars lying along the LMS-1's orbit. These RR Lyraes are also shown in Figure~\ref{fig:Fig_compare_phase-space}, where we compare their distances with the simulation of LMS-1. Note that some of these RR Lyrae could be field contaminants. However, it is hard to numerically gauge the level of contamination. Therefore, in order to ascertain the statistical significance of the RR Lyrae sample found along the LMS-1's orbit, we follow a pragmatic approach. To this end, we follow exactly the same RR Lyrae selection procedure as above, except this time inverting the orbit model in the velocity space (i.e., setting $\mathbf{v}\to \mathbf{-v}$). In this case fewer RR Lyrae are selected, $N=91$, implying that the original detection along LMS-1's orbit has a statistical significance of $2.9\sigma$. 

We conclude that, in spite of the spatially variable incompleteness of the available RR Lyrae sample, a marginally significant overdensity is detected, suggesting that LMS-1 does contain RR Lyrae. However, this conclusion should be confirmed with the new RR Lyrae catalogs that will be available in future \Gaia DRs.

\section{Dynamical model of LMS-1}\label{sec:Nbody_model}

Dynamical modelling of a stellar stream allows us to understand the process of tidal stripping of its progenitor, the overall formation and evolution of the stream itself, and the extent to which the member stars could be spread out in phase-space of the Galaxy. 

We try $6$ dynamical models of LMS-1 as follows. We use the best-fit orbit model of LMS-1 and integrate it backward in time for $T \sim 8\Gyr$ and $T \sim 3\Gyr$, and the final values of these orbits provide the starting phase-space positions to launch the progenitors forward in time. The reason for trying these two values is to examine how the resulting stream models differ if the accretion was early or recent. At each of these two starting phase-space locations, we further try  three progenitor models using Plummer spheres of masses $M_{\rm halo}=[1\times10^8,10\times 10^8,100\times 10^8]\msun$ with the corresponding scale radii  $r_s=[1.0,2.0,2.5]\kpc$, respectively. These three mass values are motivated by the [Fe/H] value of LMS-1 (see Section~\ref{sec:Discussion_Conclusion}), and they allow us to examine the differences in a low-mass and a high-mass stream model. Note that there is no additional stellar component in these models, so the particles should be interpreted as dark matter. We also account for the self-gravity of the progenitor as stream particles are released. To initialise the progenitor model and to run the simulation, we use the \texttt{gala} package \citep{Price-Whelan_GALA_2017}. This system is evolved in the \texttt{MilkyWayPotential} Galactic mass model \citep{Price-Whelan_GALA_2017}, which is a static, axisymmetric model of the Milky Way. This model is similar to the one used above, and also has a very similar mass profile (at least out to the Galactocentric distance of $50\kpc$).

After running these $6$ simulations, we visually compared the resulting stream models with the observed phase-space data of LMS-1. The model $[M_{\rm halo},T]=[1\times10^8\msun, 3\Gyr]$ was least probable as it possessed very low phase-space dispersion and failed to reproduce the measured $\sigma_{\rm v_T \, int}$ and $\sigma_{\rm vlos, int}$ of LMS-1. The model $[M_{\rm halo},T]=[1\times10^8\msun, 8\Gyr]$ possessed higher phase-space dispersion, that was at least qualitatively consistent with that of LMS-1. This model implies that the observed segment of LMS-1 is comprised of multiple wrappings of this stream. However, this model was not spread out enough in phase-space so as to be able to reproduce the phase-space location of NGC~5024. The model $[M_{\rm halo},T]=[10\times10^8\msun, 3\Gyr]$ also failed to reproduce the phase-space location of NGC~5024. As for the high mass models corresponding to $M_{\rm halo}=100\times10^8\msun$, they failed to reproduce the densest (north) part of LMS-1 with a high contrast, and the overall phase-space structure of the simulated stream did not match very well with the observed data. Finally, we found that the model $[M_{\rm halo},T]=[10\times10^8\msun, 8\Gyr]$ was the best of the set as it matched quite well with the LMS-1 stars in  position, proper motions, and line-of-sight velocities, and was able to reproduce the phase-space locations of NGC~5024 and NGC~5053 (as shown in Figure~\ref{fig:Fig_compare_phase-space}). This model comprises multiple wrappings, and it further suggests that the detected LMS-1 is the trailing arm of the overall stream. However, this inference depends on the present-day phase-space location of the (completely dissolved) LMS-1 progenitor, that we assumed to be the mid-point of the LMS-1 stream.

Figure~\ref{fig:Fig_compare_phase-space} also compares the phase-space measurements of the sub-structures sub-1, sub-2, sub-3, and the southern agglomeration of stars with the best model described above. The qualitative match between the positions and proper motions of the southern stars with the simulations suggests that LMS-1 is likely completely wrapped around the inner Milky Way halo. If LMS-1 and the southern stars are truly connected, then using the model we predict for the southern stars that their $v_{\rm los}$ should lie in the range $\sim -50$ to $150\kms$. The features sub-1, sub-2 and sub-3 overlap with the simulation in position and proper motion space. The model further suggests that sub-3 is part of the leading arm of the overall stream. However, at this stage, it is hard to analyze further their true nature and their possible association with LMS-1. Either these features are simply the over-densities of stars along the LMS-1 stream (that \texttt{STREAMFINDER} detects as independent stream-like features), or they are portions of LMS-1 that are wrapped by $\sim360\deg$ (or multiples thereof). This model is also able to reproduce the distances of the RR Lyraes in the southern sky, that also implies that most of the detected RR Lyrae lie in the leading arm of the stream. 

Figure~\ref{fig:Fig_compare_phase-space} also shows some additional objects that overlap with LMS-1's simulation in phase-space. These objects include NGC~5024, NGC~5053 and Pal~5 (data taken from \citealt{Baumgardt2019}) and the Indus stream (data taken from \citetalias{Ibata_GaiaEDR3} and \citealt{Ji_S5_2020}). Pal~5 overlaps with the leading arm of the simulation. The phase-space measurements of Indus matches with the southern segment in the position and the proper motion space, and also with RR Lyrae in the distance space. The reason for particularly highlighting Pal~5 and Indus is that we found these objects to strongly overlap with LMS-1 in the dynamical action-energy space (see Section~\ref{sec:ActionEnergy}).

\begin{table*}
\centering
\caption{Action, energy, other orbital properties and metallicity of the LMS-1 stream and of the possibly associated objects.}
\label{tab:table_orbits}
\begin{tabular}{|l|l|l|l|l|l|l|l|}

\hline
\hline
{\bf Name} & ($J_R,J_\phi,J_z$) & Energy & $r_{\rm peri}$ & $r_{\rm apo}$ & $z_{\rm max}$ & eccentricity & [Fe/H] \\
& [$\kpc \kms$] & [$\kmmss$] & [$\kpc$] & [$\kpc$] & [$\kpc$]  & &[dex]\\
\hline
\hline
&&&&&&&\\

{\bf LMS-1} (simulation) & $( 107 ^{+ 153 }_{- 66 }, -635 ^{+ 273 }_{- 390 }, 1907 ^{+ 1433 }_{- 852 })$ & $ -130836 ^{+ 19187 }_{- 27298 }$ & $ 9.2 ^{+ 6.1 }_{- 3.3 }$ & $ 16.5 ^{+ 8.2 }_{- 6.7 }$ & $ 15.8 ^{+ 8.3 }_{- 6.5 }$ & $ 0.24 ^{+ 0.11 }_{- 0.09 }$ & $-2.1\pm0.4$\\

&&&&&&&\\

{ \bf NGC~5024} & $( 473 ^{+ 10 }_{- 10 }, -688 ^{+ 10 }_{- 9 }, 2048 ^{+ 23 }_{- 21 })$ & $ -124015 ^{+ 280 }_{- 299 }$ & $ 8.4 ^{+ 0.1 }_{- 0.1 }$ & $ 21.9 ^{+ 0.1 }_{- 0.1 }$ & $ 21.1 ^{+ 0.1 }_{- 0.1 }$ & $ 0.44 ^{+ 0.01 }_{- 0.01 }$ & $-2.07\pm0.07$\\
&&&&&&&\\

{ \bf NGC~5053} & $( 192 ^{+ 9 }_{- 9 }, -625 ^{+ 9 }_{- 9 }, 2103 ^{+ 22 }_{- 22 })$ & $ -130167 ^{+ 261 }_{- 264 }$ & $ 9.5 ^{+ 0.2 }_{- 0.2 }$ & $ 17.6 ^{+ 0.0 }_{- 0.0 }$ & $ 17.1 ^{+ 0.0 }_{- 0.0 }$ & $ 0.3 ^{+ 0.01 }_{- 0.01 }$ & $-2.45\pm0.04$\\
&&&&&&&\\

{ \bf Pal~5} & $( 193 ^{+ 17 }_{- 17 }, -1052 ^{+ 24 }_{- 26 }, 1634 ^{+ 37 }_{- 33 })$ & $ -130902 ^{+ 590 }_{- 537 }$ & $ 9.3 ^{+ 0.3 }_{- 0.3 }$ & $ 17.4 ^{+ 0.0 }_{- 0.0 }$ & $ 15.9 ^{+ 0.0 }_{- 0.0 }$ & $ 0.3 ^{+ 0.02 }_{- 0.01 }$ & $-1.56\pm 0.1$\\

&&&&&&&\\

{ \bf Indus} & $( 58 ^{+ 58 }_{- 36 }, -996 ^{+ 144 }_{- 123 }, 2095 ^{+ 171 }_{- 178 })$ & $ -127649 ^{+ 5352 }_{- 5778 }$ & $ 12.1 ^{+ 0.5 }_{- 0.6 }$ & $ 16.7 ^{+ 2.8 }_{- 2.6 }$ & $ 15.8 ^{+ 2.6 }_{- 2.3 }$ & $ 0.16 ^{+ 0.05 }_{- 0.06 }$
 & $-2$\\
 

&&&&&&&\\

\hline
\hline

\end{tabular}
\end{table*}
\begin{figure*}
\begin{center}
\vspace{-0.3cm}
\includegraphics[width=0.8\hsize]{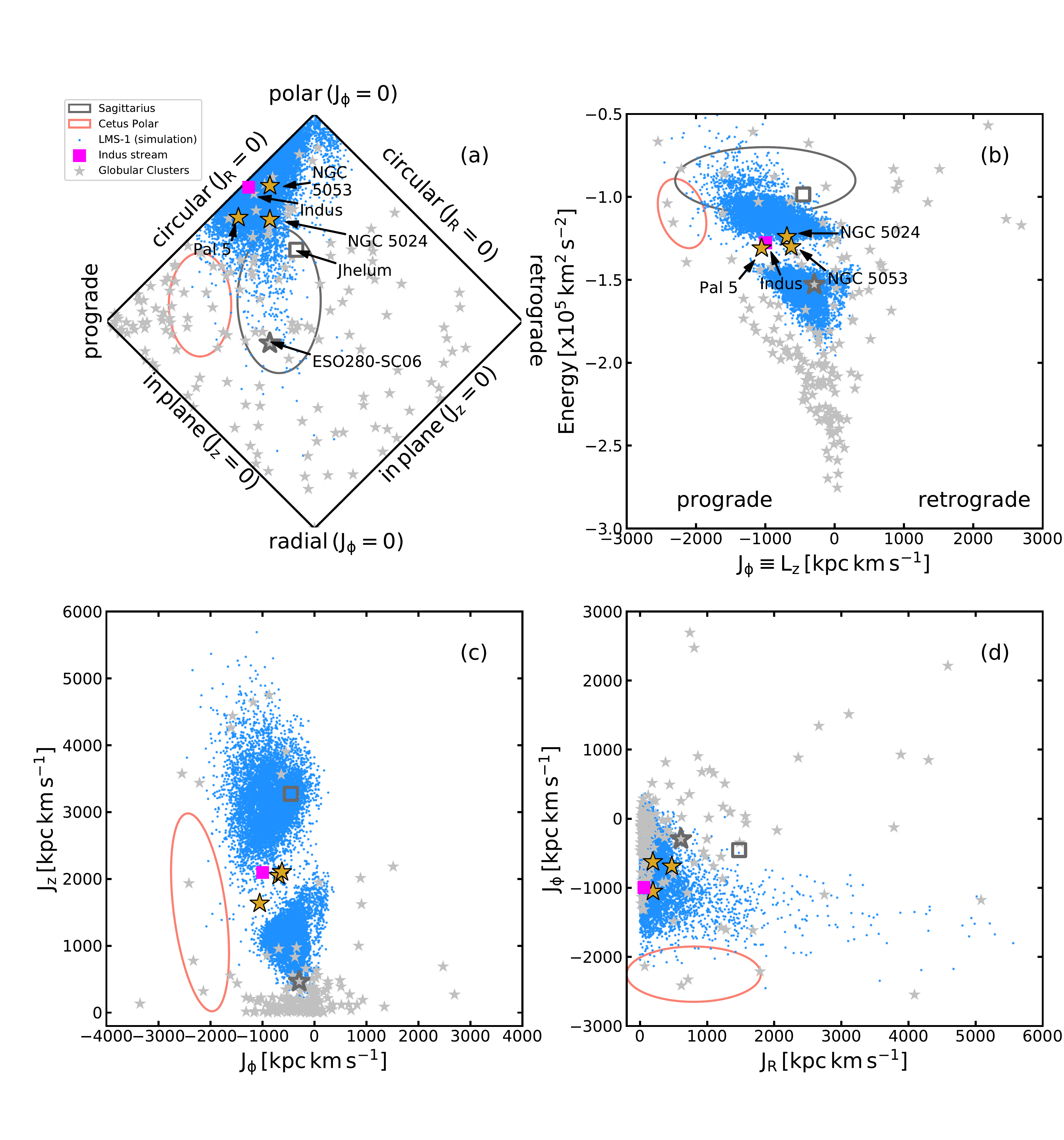}
\end{center}
\vspace{-1.2cm}
\caption{Comparing the action ($\mathbf{J}$) and energy ($E$) of LMS-1 with the globular clusters and streams of the Milky Way. (a)  The projected action-space map. The horizontal axis is $J_\phi/J_{\rm total}$ and the vertical axis is $(J_z - J_R)/J_{\rm total}$, where $J_{\rm total}=J_R+J_z+|J_\phi|$. The blue points represent the LMS-1 simulation, the golden `star' markers represent the globular clusters that have similar actions and energy as that of LMS-1, the gray `star' markers represent globular clusters of the Milky Way, and the pink `square' marker represents the Indus stream that is also strongly co-incident with LMS-1 in action-energy space. The unfilled `square' and `star' markers correspond to the ``Jhelum'' stream and ESO280-SC06 cluster, respectively (see text). The panel also highlights the approximate locations of the Sagittarius stream and the Cetus stream. (b) The z-component of angular momentum ($J_\phi$) vs. the orbital energy. The gap in the LMS-1 simulation points marks the location of the progenitor, and the two tidal arms are located on opposite sides of the gap. (c) $J_\phi$ vs. $J_z$. (d) $J_R$ vs. $J_\phi$.}
\label{fig:Fig_compare_action_energy}
\end{figure*}
\section{Objects associated with LMS-1 in action-energy space}\label{sec:ActionEnergy}

In Sections~\ref{sec:Data} and \ref{sec:orbit} we showed that both the [Fe/H] and the orbital properties of globular clusters NGC~5053 and NGC~5024 are very similar to that of the LMS-1 stream. This already provides strong evidence that these objects were accreted inside the parent LMS-1 galaxy. These associations also motivated us to examine whether there exist other Galactic globular clusters and streams that might also be associated with LMS-1. 

To search for all the objects that possibly accreted within LMS-1's parent galaxy, a strategy is to find those objects that cluster close to LMS-1 in energy-action ($E,\mathbf{J}$) space. This is because orbital parameters $E$ and $\mathbf{J}$ are preserved for adiabatic systems. This implies that objects that accrete inside the same parent galaxy, that naturally have very similar orbits, should remain tightly clustered in ($E,\mathbf{J}$); even long after they have been tidally removed from the progenitor. Conceptually, actions represent the amplitude of the orbit along different directions in a given coordinate system. We analyse the actions in cylindrical coordinates, i.e., in the $\mathbf{J}\equiv(J_R,J_\phi,J_z)$ system, where $J_\phi$ corresponds to the z-component of angular momentum. 

Figure~\ref{fig:Fig_compare_action_energy} shows the derived $E$ and $\mathbf{J}$ values for the LMS-1 stream, the Indus stream and all the globular clusters of the Milky Way. For LMS-1, these values are computed using the phase-space positions of the simulated particles that gives an ellipsoidal distribution for LMS-1 in ($E,\mathbf{J}$) space. For the globular clusters, we compute these values using their phase-space measurements from \cite{Baumgardt2019}. Moreover, for every globular cluster, we sample $N=1000$ orbits using the observational uncertainties. Here, the ($E,\mathbf{J}$) value corresponds to their mean values. To compute these values for Indus, we fit an orbit to this stream following a similar procedure as described in Section~\ref{sec:orbit}. The positions, parallaxes and proper motions of the Indus stars, along with the observational uncertainties, are taken from \citetalias{Ibata_GaiaEDR3} and the line-of-sight velocities are taken from \cite{Ji_S5_2020}. Here, the ($E,\mathbf{J}$) value corresponds to the best fit orbit model of Indus. We tried a few other prograde streams as well (e.g., Sylgr, Jhelum) but found that they do not strongly overlap with LMS-1 in this dynamical space. As an example we show Jhelum in Figure~\ref{fig:Fig_compare_action_energy} that lies beyond the LMS-1 distribution. Note that since Jhelum is unrelated to LMS-1, it effectively implies that it is also unrelated to Indus. This result is different from a recent analysis that indicates a common origin of the Indus and Jhelum streams \citep{Bonaca2021}. This conclusion is also different from the common assumption in the literature that Indus and Jhelum have very similar orbital properties, and thus represent different orbital wraps of the very same stream (e.g., see \citealt{Shipp2018, Bonaca_2019}; although their assumption is based on analyses that included either zero or only partial velocity information on these streams).

As is evident in Figure~\ref{fig:Fig_compare_action_energy}, NGC~5024, NGC~5053, Pal~5 and Indus are remarkably co-incident with LMS-1 in the ($E,\mathbf{J}$) space (also see Table~\ref{tab:table_orbits}). From this strong dynamical clustering, it is tempting to instantly conclude that all of these objects were accreted inside the LMS-1's progenitor galaxy. However, despite Pal~5's striking similarity with LMS-1 in ($E,\mathbf{J}$), their association is not clear. This is because 1) Pal~5 has a slightly higher metallicity than LMS-1 (see Figure~\ref{fig:Fig_FeH}), and 2) the orbital pole of Pal~5 is almost opposite to that of LMS-1, $\theta \sim 143\deg$ apart. This difference in their orbital poles is evaluated as $\rm{\theta=cos^{-1} (\hat{L_1}.\hat{L_2})}$; where $\rm{\hat{L_1},\hat{L_2}}$ refers to the present day unit vector angular momenta of LMS-1 and Pal~5, respectively. In other words, at the present day, Pal~5 orbits in the opposite direction in the Y-Z plane compared to the orbit of LMS-1 (i.e., the x-component of angular momentum has an opposite sign). These differences render the association of Pal~5 with LMS-1 a bit doubtful, albeit worthy of further investigation. Moreover, unlike \cite{Naidu2020}, we find that the globular cluster ESO 280-SC06 lies beyond the dynamical region of LMS-1 (as shown in Figure~\ref{fig:Fig_compare_action_energy}), and is not associated with this dwarf galaxy stream. Also, the orbital pole of ESO 280-SC06 is $\theta \sim 84\deg$ apart from the orbital pole of LMS-1. In conclusion, the similarity of NGC~5024, NGC~5053 and Indus with LMS-1 in ($E,\mathbf{J}$) space, and also in [Fe/H], strongly indicates that these objects were accreted inside the LMS-1's parent galaxy.

The above result is based on the simulation where the progenitor accretes along the best-fit orbit of LMS-1 (obtained in Section~\ref{sec:orbit}). Instead, if we assume NGC~5053 (or NGC~5024) as the nuclear star cluster, and run the LMS-1 simulation using its orbit, we obtain a slightly different result. Particularly, in this new scenario, Pal~5 does not strongly overlap with LMS-1 in ($E,\mathbf{J}$) space. However, another globular cluster can now be associated with LMS-1, namely NGC~6715 (a.k.a. M~54, the nuclear star cluster of the Sagittarius dwarf). This examination makes us confident that atleast NGC~5053, NGC~5024 and Indus are associated with LMS-1. However, the fact that we find M~54 to overlap with LMS-1 in ($E,\mathbf{J}$) space indicates that the association of different objects purely based on the dynamical parameters can be misleading, and therefore one should be extremely cautious during such analyses. This is the reason that along with comparing the dynamical properties of LMS-1 and its associated objects, we also compare their [Fe/H] (see Section~\ref{subsec:FeH_LMS1}) and stellar population (see Section~\ref{sec:Discussion_Conclusion}).

\section{Discussion and Conclusion}\label{sec:Discussion_Conclusion}

We have presented the first comprehensive study of the LMS-1 stellar stream, based on the astrometric dataset of \Gaia\ EDR3 and spectroscopic datasets of LAMOST, SDSS/SEGUE and APOGEE. This stream was detected by the \texttt{STREAMFINDER} algorithm during our hunt for wide streams in the \Gaia\ EDR3 dataset. We found that LMS-1 has a broad physical width ($\sim700\pc$), a very low metallicity ($\rm{\langle [Fe/H] \rangle}\sim-2.1$) with significant metallicity dispersion ($\sigma_{\rm [Fe/H]}\sim 0.4$), and these inferences together imply that this stream was produced from a dwarf galaxy. However, we do note that no progenitor of LMS-1 has so far been detected, likely because it has been completely disrupted. Below, we briefly discuss some of the properties of LMS-1 that make it a very interesting dwarf galaxy stream. 

LMS-1's orbit is very circular (eccentricity$\sim0.2$) and possesses a strikingly small apocentre ($\sim16\kpc$). These orbital properties make LMS-1 a very intriguing structure because dwarf galaxy streams, and even `intact' dwarf galaxies, generally lie on highly eccentric orbits (since accretions occur radially to the host galaxy) and they also possess quite large apocentres. For instance, the Orphan stream has an eccentricity of $\sim 0.7$ \citep{Newberg_orphan_2010}. Also, most of the dwarf galaxy streams have apocentres $\simgt35\kpc$ (this lower limit of the apogalacticon value is set by the Cetus stream, \citealt{Yam_2013}). One scenario that explains these surprising orbital properties of LMS-1 is if its parent galaxy was very massive and was accreted at a sufficiently early time to have experienced dynamical friction that eventually circularised its orbit \citep{Chandrasekhar1943}. LMS-1's orbit around the Milky Way is also quite polar, with an inclination of $\sim75\deg$ to the Galactic plane (similar to that of the Sagittarius stream) which may make it a powerful probe of the flattening of the dark matter halo in the inner regions of our Galaxy. 

LMS-1 can be observed in Figure~\ref{fig:Fig_SF_map} as a simple $60\deg$-long broad stream located towards the north of the Galactic bulge, at a distance of $\sim 20\kpc$ from the Sun. However, this stream is probably a much more complex system, completely wrapping around the inner Milky Way halo, as suggested by two main evidences: (1) the presence of multiple off-stream sub-structures adjacent to LMS-1 (namely sub-1, sub-2 and sub-3) and the agglomeration of stars in the southern galactic sky, and (2) the very large velocity dispersion of LMS-1, as measured along both the line-of-sight direction ($\sigma_{\rm v_{los}}\sim20\kms$) and the tangential direction ($\sigma_{\rm v_{T}}\sim35\kms$). Also, the CMDs of these sub-structures are plausibly similar to that of LMS-1 (see Figure~\ref{fig:Fig_CMD_LMS1_comps}), indicating similarity in their stellar population (although we note that it would be useful to obtain deeper CMDs of the sub-structures to better compare their populations). Furthermore, the positions and proper motions of these sub-structures were (qualitatively) reproducible in our N-body model (see Figure~\ref{fig:Fig_compare_phase-space}). The N-body model also predicts that if the southern component is a part of LMS-1, then its line-of-sight velocity should range between $v_{\rm los}\sim -50$ to $150\kms$. Note that we lacked the measurements of [Fe/H] and line-of-sight velocities of these sub-structures that would have been very useful to elucidate their origin and possible link with LMS-1.

\begin{figure*}
\begin{center}
\vspace{-0.3cm}
\includegraphics[width=\hsize]{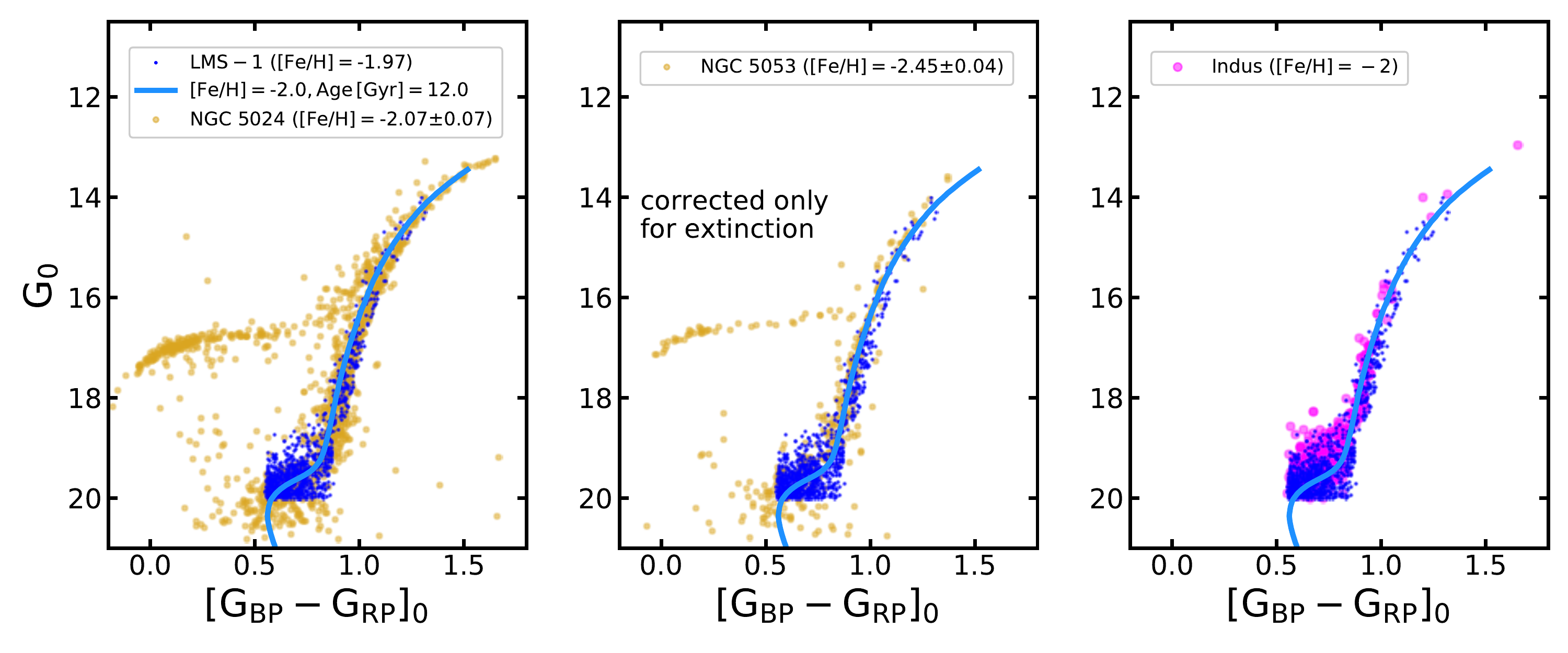}
\end{center}
\vspace{-0.5cm}
\caption{Comparing the color-magnitude distribution of LMS-1 with that of NGC~5024, NGC~5053 and Indus. All panels show photometry corrected only for extinction. The stellar population model (blue line) has been shifted to account for the heliocentric distance of LMS-1.}
\label{fig:Fig_CMD}
\end{figure*}

Therefore, in light of these limited observations, we consider three scenarios and deem that their combined effect could explain both the presence of the sub-structures (proximal to LMS-1's orbit) and LMS-1's large velocity dispersion. (A) {\it LMS-1 is highly phase-mixed:} Given that LMS-1 orbits the inner regions of the Milky Way halo, it is possible that this system accreted early ($\simgt8-10\Gyr$ ago), was severely damaged by the large tidal forces of the inner Galactic potential, and the eventual phase-mixing increased the velocity dispersion of the stream. (B) {\it LMS-1 and the observed sub-structures are orbital wraps of the same stream:} A dwarf galaxy on a similar orbit as that of LMS-1 would completely wrap around the inner halo in $\sim 0.5\Gyr$ (as suggested by our controlled simulations). Therefore, it is possible that the observed segment of LMS-1 is actually a projection of multiple wrappings of the stream along the line-of-sight direction, and we have ended up measuring their cumulative velocity dispersion. The same scenario also suggests that the observed sub-structures correspond to highly-dense segments of LMS-1 that are wrapped by $\sim360\deg$ (or multiples thereof). (C) {\it Precession of LMS-1:} The orbital planes continuously evolve in flattened potentials \citep{Erkal_2016}. If the Milky Way's dark matter halo is flattened in the inner regions (which most likely is the case due to the presence of the Galactic disk), then the angular momentum vector of LMS-1 has been continuously evolving during its lifetime, and this precession effect could be the source of its large velocity dispersion. Moreover, due to precession, streams lying along nearly polar orbits are expected to quickly spread in phase-space \citep{Erkal_2016}. This effect may also explain the estimated broad physical width and large velocity dispersion of LMS-1.

Future observational and modeling studies will be required to examine the association of these sub-structures with LMS-1, and to test whether LMS-1 completely wraps around the inner Milky Way halo. For this, the prime requirement will be to obtain measurements of [Fe/H] and line-of-sight velocities of these sub-structures. A similarity in [Fe/H] (and element abundances) between these sub-structures and LMS-1 would strengthen the case for their association. Velocity measurements will also be useful to infer whether these sub-structures can be reproduced in detailed N-body simulations, alongside LMS-1 itself. For this, it will be important to model both the parent LMS-1 galaxy and the Galactic potential more realistically to properly consider both the precession effect and the dynamical friction effect in the inner regions of the Milky Way.

We also found that in action-energy ($E,\mathbf{J}$), and other orbital properties, LMS-1 is remarkably similar to the globular clusters NGC~5024, NGC~5053, Pal~5 and the stellar stream ``Indus''. The dynamical association of these objects is shown in Figure~\ref{fig:Fig_compare_action_energy} (also see Table~\ref{tab:table_orbits}). The co-incidence of LMS-1’s orbit with that of NGC~5024 and NGC~5053 was also proposed in \cite{Yuan2019_CPS, Naidu2020}, and here we have confirmed this through proper orbital analysis. Among these objects, NGC~5024, NGC~5053 and Indus also have very similar orbital poles (at the present day), as well as [Fe/H], as that of LMS-1. On the other hand, Pal~5's orbital pole is $\sim 140\deg$ different from LMS-1, and it is also slightly more metal rich than LMS-1 (see Figure~\ref{fig:Fig_FeH}), casting doubt on an association.  Furthermore, the CMDs of NGC~5053, NGC~5024 and Indus match well with that of LMS-1 (see Figure~\ref{fig:Fig_CMD}), implying similarity also in their stellar populations. In summary, the remarkable similarity of LMS-1, NGC~5053, NGC~5024 and Indus in their dynamics, [Fe/H] and stellar population indicates that all of these objects were accreted inside the same parent galaxy.

Given that Indus is produced from a dwarf galaxy\footnote{Indus has recently been interpreted as a low-mass dwarf galaxy stream, and not a globular cluster stream, based on chemical signatures of its stars \citep{Ji_S5_2020}.}, this further implies that Indus's progenitor was a satellite dwarf galaxy of the parent LMS-1 dwarf. Another possibility is that Indus is not an independent object, but simply a subpart of the southern wrap of LMS-1 (see Figure~\ref{fig:Fig_compare_phase-space}a). However, this scenario holds less merit because Indus is a narrow and highly-coherent stream that is clearly distinguishable from the broad and diffuse southern wrap of LMS-1 (see Figure~\ref{fig:Fig_compare_phase-space}f). This indicates that the progenitor of Indus must have been an independent object that was smaller in size, and much denser, than the parent LMS-1 dwarf. This scenario is also supported by cosmological simulations that have shown that smaller galaxies formed at higher redshifts (like Indus), that eventually ended up being satellites of larger galaxies (like LMS-1), possess higher densities \citep{Bullock2001}.

\begin{figure*}
\begin{center}
\vspace{-0.3cm}
\includegraphics[width=\hsize]{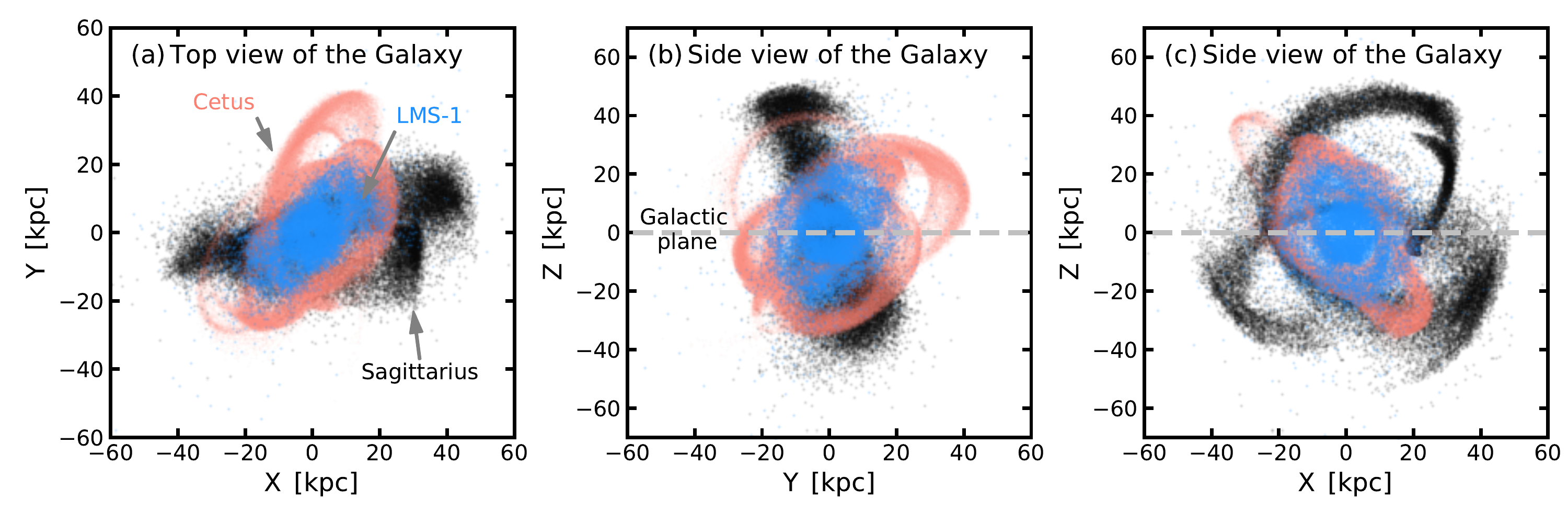}
\end{center}
\vspace{-0.5cm}
\caption{Comparing the spatial distributions of stellar stream models of LMS-1, Cetus and Sagittarius in the Galactocentric coordinate system.}
\label{fig:Fig_compare_LMS_Sgr_CPS}
\end{figure*}
\begin{figure}
\begin{center}
\vspace{-0.3cm}
\includegraphics[width=\hsize]{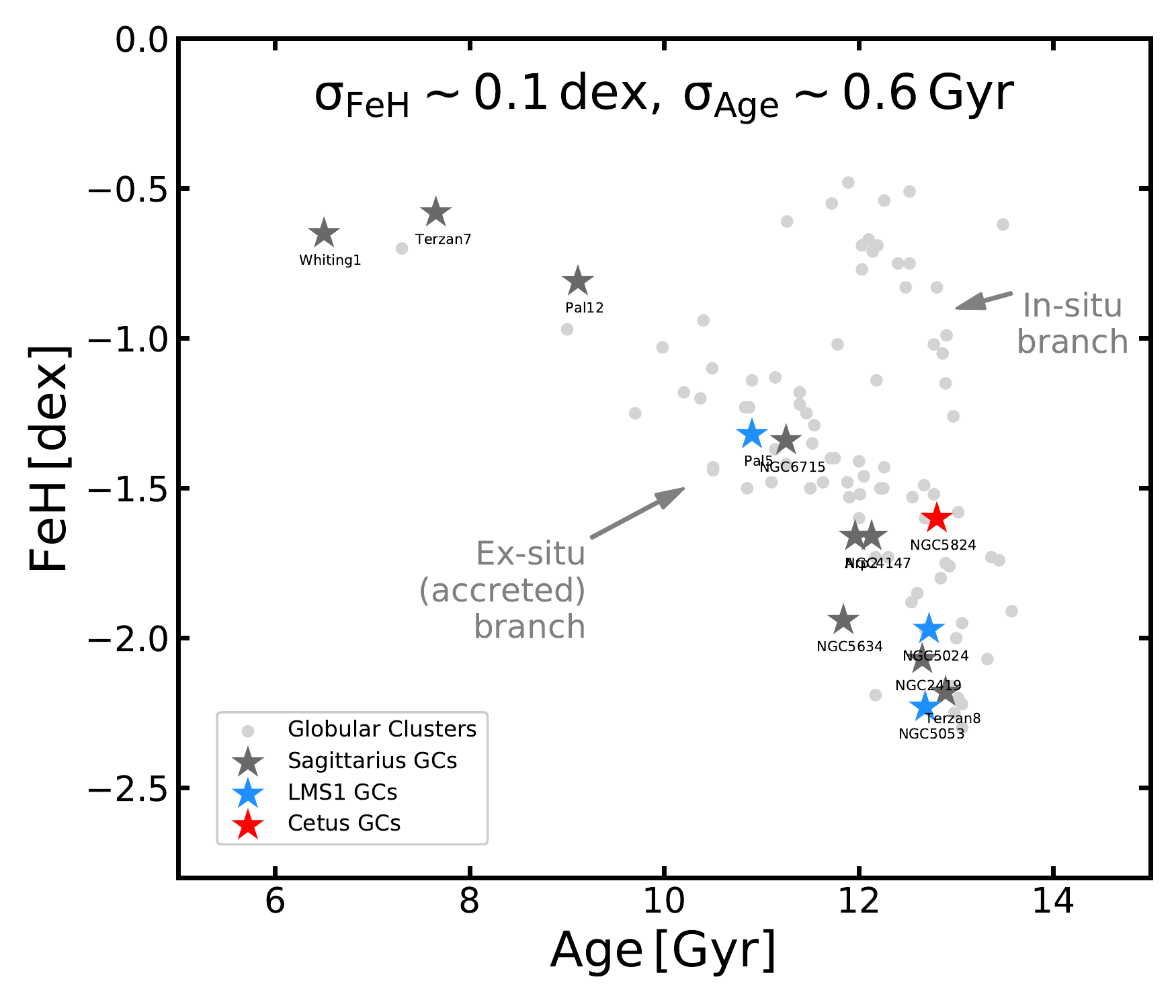}
\end{center}
\vspace{-0.5cm}
\caption{The age-metallicity distribution of the Milky Way globular clusters. Typical uncertainties on the parameters are mentioned on the plot. Globular clusters associated with LMS-1, Cetus and Sagittarius are shown for comparison.}
\label{fig:Fig_compare_GCs_LMS_Sgr_CPS}
\end{figure}

We also found that the orbit of NGC~5053 is strikingly similar to that of LMS-1 (see Figure~\ref{fig:Fig_Orbit_fit}). At some level, this indicates that NGC~5053 probably originated from the very central regions of LMS-1's parent galaxy; reminiscent of the M~54 star cluster that currently resides inside the very nucleus of the Sagittarius dwarf galaxy (e.g., \citealt{Bellazzini_2008}). This particular inference is different from that of \cite{Yuan2020} who otherwise proposed that it is NGC~5024 that originated from the nucleus of LMS-1's progenitor (based only on the fact that NGC~5024 is more massive than NGC~5053). However, we note that neither NGC~5024 nor NGC~5053 make good candidates for nuclear star clusters because neither of them are Type II clusters\footnote{Type II clusters are those that exhibit complex/multiple stellar population, and where first and second generations of stars are distinguishable. Examples of Type II clusters include M~54, $\omega$-Centauri, that have been confirmed to have formed in the central regions of their parent dwarfs (e.g., \citealt{Pfeffer2021}). On the other hand, Type I clusters correspond to those that exhibit simpler stellar population, and where first and second generations of stars are not easily distinguishable (likely because they host only one generation of stars).} \citep{Milone2017} and they also do not show any [Fe/H] spread \citep{Boberg_NGC5053_2015, Boberg_NGC5024_2016}. Nonetheless, given that the difference in the orbital energies of NGC~5053 and LMS-1 is smaller than that between NGC~5024 and LMS-1 suggests that NGC~5053 must have formed closer to the LMS-1 dwarf's centre.

We now predict LMS-1 parent galaxy's mass given that LMS-1's mean metallicity is ${\rm \langle [Fe/H] \rangle} \sim -2.1$ and it brought in no less than two globular clusters. The [Fe/H] measurement of LMS-1 implies that the stellar mass of its parent dwarf galaxy was $M_{*}\sim10^5\msun$ (from the present-day Stellar Mass-Stellar Metallicity Relation of dwarf galaxies, \citealt{Kirby2013}), and this further indicates that the dark matter halo mass of the progenitor was $M_{\rm halo}\sim 10^9\msun$ (from the stellar-to-halo-mass relation from \citealt{Read2017}). These estimates suggest that the parent LMS-1 dwarf was probably similar to the Draco dwarf galaxy that has ${\rm \langle [Fe/H] \rangle}= -1.93$ \citep{Kirby2011} and $M_{*}\sim 3\times 10^5\msun$ \citep{McConnachie2012}. However, this estimate implies that the stellar mass of the parent LMS-1 galaxy was lower than the combined mass of the two clusters: NGC~5024 ($M\sim5\times10^5\msun$) and NGC~5053 ($M\sim0.5\times10^5\msun$). This in stark contrast with the observations of the present day dwarfs, where the ratio between the total mass of their member clusters and the stellar mass of the host dwarf is measured to be much smaller than unity. For instance, in Fornax dwarf, this mass fraction is $\sim 0.2$ at [Fe/H]$\simlt-2.1$ \citep{de_Boer_2016}. We argue that $M_{*}$-[Fe/H] relation can only provide a lower limit on the $M_{*}$ value of LMS-1-like dwarfs that accreted very early in time, since this relation possesses a redshift dependence \citep{Torrey_2019} and, in general, also possesses a significant scatter.

An alternate way to estimate $M_{*}$ is to first constrain the value of $M_{\rm halo}$. $M_{\rm halo}$ of a parent galaxy can be constrained from the total mass of the member globular clusters \citep{Hudson2014}. Adopting the masses of NGC~5053, NGC~5024 from \cite{Baumgardt2019}, we obtain $M_{\rm halo}\sim10^{9-10}\msun$. This halo mass implies a stellar mass of $M_{*}\sim 10^{6-7}\msun$. Note that in this case, the stellar mass is larger than the combined masses of globular clusters. Overall, these properties suggest that the parent LMS-1 dwarf was similar to the Fornax dwarf galaxy (that hosts a population of at least $5$ globular clusters and has $M_{*}\sim 2\times 10^7\msun$, \citealt{McConnachie2012}). These mass estimates of LMS-1's parent galaxy justify the assumptions we made for our N-body models.

To put LMS-1 into some perspective, we now briefly examine how this stream compares with two other dwarf galaxy streams of the Milky Way: the Sagittarius stream and the Cetus stream. The prime reason for comparing these particular streams is that all of them lie along prograde orbits, and possess quite polar structures (inclined at angles $\simgt60\deg$ to the Galactic plane). The dynamical models of these streams are shown in Figure~\ref{fig:Fig_compare_LMS_Sgr_CPS}, where the Sagittarius model is adopted from \cite{Law2010}, the Cetus model is taken from \cite{Chang_2020_Cetus} and the LMS-1 model corresponds to the one that we found most suitable in Section~\ref{sec:Nbody_model}. Among these streams, the colossal Sagittarius spans a large range in Galactocentric distance from $15\kpc$ to $100\kpc$ \citep{Belokurov2014}, and is a major contributor to the stellar populations of the outer Galactic halo \citep{Newberg2002, Belokurov2006_Sgr}. This stream has also contributed $6-8$ globular clusters to the Milky Way \citep{Massari2019, Bellazzini2020}, with M~54 being the nuclear star cluster of the parent Sagittarius galaxy. Moreover, this massive stream has a large spread in metallicity from [Fe/H]$\sim -0.5$ to $-2.5$ \citep{Hayes_2020_Sgr}. Recent dynamical models of Sagittarius indicate that it is a dynamically young system that recently accreted into our Galaxy, $\sim6\Gyr$ ago \citep{Ruiz-Lara2020, Lian2020}. The next stream that we consider is Cetus that orbits intermediate Galactocentric distances from $24\kpc$ and $36\kpc$ on a nearly circular trajectory \citep{Yam_2013}. Furthermore, the metallicity of Cetus is [Fe/H]$\sim-2$, implying that its progenitor was probably a low-mass dwarf galaxy (similar to that of LMS-1). Cetus has been dynamically linked with the globular cluster NGC~5824, that was probably the nuclear star cluster of its parent dwarf galaxy \citep{Yuan2019_CPS}. 

Figure~\ref{fig:Fig_compare_GCs_LMS_Sgr_CPS} shows age-metallicity distribution of $96$ globular clusters of the Milky Way, taken from compilation by \cite{Kruijssen_2019}. This plot shows two well-known branches: (1) a branch corresponding to nearly uniform old age group and higher metallicities, and (2) a branch corresponding to younger ages and lower metallicities (cf. \citealt{Forbes_2010}). The former branch is associated with in-situ GCs, and the latter is assumed to have emerged via accretion of clusters from ex-situ dwarf galaxies. We particularly highlight the globular clusters associated with Sagittarius, Cetus and LMS-1. The location of NGC~5024, NGC~5053 and Pal~5 on this map provides supporting evidence of their accreted origin.

Our analysis suggests that LMS-1 with its very small apocenter is a very interesting fossil remnant of the early formation of our Galaxy. Streams such as LMS-1 also provide a novel opportunity to study the formation and evolution of globular clusters in ancient dwarf galaxies. Future detection and analysis of other dwarf galaxy streams will allow us to build a reliable understanding of the Milky Way's assembly history, that could ultimately be compared with the predictions from the cosmological models of galaxy formation.

\section*{ACKNOWLEDGEMENTS}

We thank the referee for helpful comments and suggestions. It is a pleasure to thank Raphaël Errani for insightful conversations. KM acknowledges support  from the $\rm{Vetenskapsr\mathring{a}de}$t (Swedish Research Council) through contract No. 638-2013-8993 and the Oskar Klein Centre for Cosmoparticle Physics. RI, NM and ZY acknowledge funding from the Agence Nationale de la Recherche (ANR project ANR-18-CE31-0006, ANR-18-CE31-0017 and ANR-19-CE31-0017), from CNRS/INSU through the Programme National Galaxies et Cosmologie. RI, NM and AA acknowledge funding from the European Research Council (ERC) under the European Unions Horizon 2020 research and innovation programme (grant agreement No. 834148). MB acknowledges the financial support to this research by INAF, through the Mainstream Grant 1.05.01.86.22 assigned to the project ``Chemo-dynamics of globular clusters: the Gaia revolution'' (P.I. E. Pancino). 

This work has made use of data from the European Space Agency (ESA) mission {\it Gaia} (\url{https://www.cosmos.esa.int/gaia}), processed by the {\it Gaia} Data Processing and Analysis Consortium (DPAC,
\url{https://www.cosmos.esa.int/web/gaia/dpac/consortium}). Funding for the DPAC has been provided by national institutions, in particular the institutions participating in the {\it Gaia} Multilateral Agreement. 
 
Funding for SDSS-III has been provided by the Alfred P. Sloan Foundation, the Participating Institutions, the National Science Foundation, and the U.S. Department of Energy Office of Science. The SDSS-III web site is http://www.sdss3.org/.

SDSS-III is managed by the Astrophysical Research Consortium for the Participating Institutions of the SDSS-III Collaboration including the University of Arizona, the Brazilian Participation Group, Brookhaven National Laboratory, Carnegie Mellon University, University of Florida, the French Participation Group, the German Participation Group, Harvard University, the Instituto de Astrofisica de Canarias, the Michigan State/Notre Dame/JINA Participation Group, Johns Hopkins University, Lawrence Berkeley National Laboratory, Max Planck Institute for Astrophysics, Max Planck Institute for Extraterrestrial Physics, New Mexico State University, New York University, Ohio State University, Pennsylvania State University, University of Portsmouth, Princeton University, the Spanish Participation Group, University of Tokyo, University of Utah, Vanderbilt University, University of Virginia, University of Washington, and Yale University.

Guoshoujing Telescope (the Large Sky Area Multi-Object Fiber Spectroscopic Telescope LAMOST) is a National Major Scientific Project built by the Chinese Academy of Sciences. Funding for the project has been provided by the National Development and Reform Commission. LAMOST is operated and managed by the National Astronomical Observatories, Chinese Academy of Sciences.

\bibliographystyle{apj}
\bibliography{ref1}

\begin{thebibliography}{}
\expandafter\ifx\csname natexlab\endcsname\relax\def\natexlab#1{#1}\fi

\bibitem[{{Arentsen} {et~al.}(2019){Arentsen}, {Prugniel}, {Gonneau},
  {Lan{\c{c}}on}, {Trager}, {Peletier}, {Lyubenova}, {Chen}, {Falc{\'o}n
  Barroso}, {S{\'a}nchez Bl{\'a}zquez}, \& {Vazdekis}}]{Arentsen_2019}
{Arentsen}, A., {Prugniel}, P., {Gonneau}, A., {et~al.} 2019, \aap, 627, A138

\bibitem[{{Baumgardt} {et~al.}(2019){Baumgardt}, {Hilker}, {Sollima}, \&
  {Bellini}}]{Baumgardt2019}
{Baumgardt}, H., {Hilker}, M., {Sollima}, A., \& {Bellini}, A. 2019, \mnras,
  482, 5138

\bibitem[{{Bell} {et~al.}(2008){Bell}, {Zucker}, {Belokurov}, {Sharma},
  {Johnston}, {Bullock}, {Hogg}, {Jahnke}, {de Jong}, {Beers}, {Evans},
  {Grebel}, {Ivezi{\'c}}, {Koposov}, {Rix}, {Schneider}, {Steinmetz}, \&
  {Zolotov}}]{Bell2008}
{Bell}, E.~F., {Zucker}, D.~B., {Belokurov}, V., {et~al.} 2008, \apj, 680, 295

\bibitem[{{Bellazzini} {et~al.}(2020){Bellazzini}, {Ibata}, {Malhan}, {Martin},
  {Famaey}, \& {Thomas}}]{Bellazzini2020}
{Bellazzini}, M., {Ibata}, R., {Malhan}, K., {et~al.} 2020, \aap, 636, A107

\bibitem[{Bellazzini {et~al.}(2008)Bellazzini, Ibata, Chapman, Mackey, Monaco,
  Irwin, Martin, Lewis, \& Dalessandro}]{Bellazzini_2008}
Bellazzini, M., Ibata, R.~A., Chapman, S.~C., {et~al.} 2008, The Astronomical
  Journal, 136, 1147

\bibitem[{{Belokurov} {et~al.}(2018){Belokurov}, {Erkal}, {Evans}, {Koposov},
  \& {Deason}}]{Belokurov2018}
{Belokurov}, V., {Erkal}, D., {Evans}, N.~W., {Koposov}, S.~E., \& {Deason},
  A.~J. 2018, \mnras, 478, 611

\bibitem[{{Belokurov} {et~al.}(2006){Belokurov}, {Zucker}, {Evans}, {Gilmore},
  {Vidrih}, {Bramich}, {Newberg}, {Wyse}, {Irwin}, {Fellhauer}, {Hewett},
  {Walton}, {Wilkinson}, {Cole}, {Yanny}, {Rockosi}, {Beers}, {Bell},
  {Brinkmann}, {Ivezi{\'c}}, \& {Lupton}}]{Belokurov2006_Sgr}
{Belokurov}, V., {Zucker}, D.~B., {Evans}, N.~W., {et~al.} 2006, \apjl, 642,
  L137

\bibitem[{{Belokurov} {et~al.}(2007){Belokurov}, {Evans}, {Irwin},
  {Lynden-Bell}, {Yanny}, {Vidrih}, {Gilmore}, {Seabroke}, {Zucker},
  {Wilkinson}, {Hewett}, {Bramich}, {Fellhauer}, {Newberg}, {Wyse}, {Beers},
  {Bell}, {Barentine}, {Brinkmann}, {Cole}, {Pan}, \& {York}}]{Belokurov2007}
{Belokurov}, V., {Evans}, N.~W., {Irwin}, M.~J., {et~al.} 2007, \apj, 658, 337

\bibitem[{Belokurov {et~al.}(2013)Belokurov, Koposov, Evans, Peñarrubia,
  Irwin, Smith, Lewis, Gieles, Wilkinson, Gilmore, Olszewski, \&
  Niederste-Ostholt}]{Belokurov2014}
Belokurov, V., Koposov, S.~E., Evans, N.~W., {et~al.} 2013, Monthly Notices of
  the Royal Astronomical Society, 437, 116

\bibitem[{{Boberg} {et~al.}(2015){Boberg}, {Friel}, \&
  {Vesperini}}]{Boberg_NGC5053_2015}
{Boberg}, O.~M., {Friel}, E.~D., \& {Vesperini}, E. 2015, \apj, 804, 109

\bibitem[{{Boberg} {et~al.}(2016){Boberg}, {Friel}, \&
  {Vesperini}}]{Boberg_NGC5024_2016}
---. 2016, \apj, 824, 5

\bibitem[{{Bonaca} {et~al.}(2019){Bonaca}, {Conroy}, {Price-Whelan}, \&
  {Hogg}}]{Bonaca_2019}
{Bonaca}, A., {Conroy}, C., {Price-Whelan}, A.~M., \& {Hogg}, D.~W. 2019,
  \apjl, 881, L37

\bibitem[{{Bonaca} {et~al.}(2021){Bonaca}, {Naidu}, {Conroy}, {Caldwell},
  {Cargile}, {Han}, {Johnson}, {Kruijssen}, {Myeong}, {Speagle}, {Ting}, \&
  {Zaritsky}}]{Bonaca2021}
{Bonaca}, A., {Naidu}, R.~P., {Conroy}, C., {et~al.} 2021, \apjl, 909, L26

\bibitem[{{Bovy}(2015)}]{Bovy2015}
{Bovy}, J. 2015, \apjs, 216, 29

\bibitem[{{Bressan} {et~al.}(2012){Bressan}, {Marigo}, {Girardi}, {Salasnich},
  {Dal Cero}, {Rubele}, \& {Nanni}}]{Bressan2012}
{Bressan}, A., {Marigo}, P., {Girardi}, L., {et~al.} 2012, \mnras, 427, 127

\bibitem[{{Buder} {et~al.}(2018){Buder}, {Asplund}, {Duong}, {Kos}, {Lind},
  {Ness}, {Sharma}, {Bland-Hawthorn}, {Casey}, {de Silva}, {D'Orazi},
  {Freeman}, {Lewis}, {Lin}, {Martell}, {Schlesinger}, {Simpson}, {Zucker},
  {Zwitter}, {Amarsi}, {Anguiano}, {Carollo}, {Casagrande}, {{\v{C}}otar},
  {Cottrell}, {da Costa}, {Gao}, {Hayden}, {Horner}, {Ireland}, {Kafle},
  {Munari}, {Nataf}, {Nordlander}, {Stello}, {Ting}, {Traven}, {Watson},
  {Wittenmyer}, {Wyse}, {Yong}, {Zinn}, {{\v{Z}}erjal}, \& {Galah
  Collaboration}}]{buder18}
{Buder}, S., {Asplund}, M., {Duong}, L., {et~al.} 2018, \mnras, 478, 4513

\bibitem[{{Bullock} \& {Johnston}(2005)}]{Bullock2005}
{Bullock}, J.~S., \& {Johnston}, K.~V. 2005, \apj, 635, 931

\bibitem[{{Bullock} {et~al.}(2001){Bullock}, {Kolatt}, {Sigad}, {Somerville},
  {Kravtsov}, {Klypin}, {Primack}, \& {Dekel}}]{Bullock2001}
{Bullock}, J.~S., {Kolatt}, T.~S., {Sigad}, Y., {et~al.} 2001, \mnras, 321, 559

\bibitem[{{Cargile} {et~al.}(2020){Cargile}, {Conroy}, {Johnson}, {Ting},
  {Bonaca}, {Dotter}, \& {Speagle}}]{cargile20}
{Cargile}, P.~A., {Conroy}, C., {Johnson}, B.~D., {et~al.} 2020, \apj, 900, 28

\bibitem[{{Chandrasekhar}(1943)}]{Chandrasekhar1943}
{Chandrasekhar}, S. 1943, \apj, 97, 255

\bibitem[{{Chang} {et~al.}(2020){Chang}, {Yuan}, {Xue}, {Simion}, {Kang}, {Li},
  {Zhao}, \& {Zhao}}]{Chang_2020_Cetus}
{Chang}, J., {Yuan}, Z., {Xue}, X.-X., {et~al.} 2020, \apj, 905, 100

\bibitem[{{Chiba} \& {Beers}(2000)}]{Chiba2000}
{Chiba}, M., \& {Beers}, T.~C. 2000, \aj, 119, 2843

\bibitem[{{Clementini} {et~al.}(2019){Clementini}, {Ripepi}, {Molinaro},
  {Garofalo}, {Muraveva}, {Rimoldini}, {Guy}, {Jevardat de Fombelle},
  {Nienartowicz}, {Marchal}, {Audard}, {Holl}, {Leccia}, {Marconi}, {Musella},
  {Mowlavi}, {Lecoeur-Taibi}, {Eyer}, {De Ridder}, {Regibo}, {Sarro},
  {Szabados}, {Evans}, \& {Riello}}]{Clementini2019}
{Clementini}, G., {Ripepi}, V., {Molinaro}, R., {et~al.} 2019, \aap, 622, A60

\bibitem[{{Conroy} {et~al.}(2019){Conroy}, {Bonaca}, {Cargile}, {Johnson},
  {Caldwell}, {Naidu}, {Zaritsky}, {Fabricant}, {Moran}, {Rhee},
  {Szentgyorgyi}, {Berlind}, {Calkins}, {Kattner}, \& {Ly}}]{Conroy_H3_2019}
{Conroy}, C., {Bonaca}, A., {Cargile}, P., {et~al.} 2019, \apj, 883, 107

\bibitem[{{de Boer} \& {Fraser}(2016)}]{de_Boer_2016}
{de Boer}, T.~J.~L., \& {Fraser}, M. 2016, \aap, 590, A35

\bibitem[{{Dehnen} \& {Binney}(1998)}]{DehnenBin1998}
{Dehnen}, W., \& {Binney}, J. 1998, \mnras, 294, 429

\bibitem[{Erkal {et~al.}(2016)Erkal, Sanders, \& Belokurov}]{Erkal_2016}
Erkal, D., Sanders, J.~L., \& Belokurov, V. 2016, Monthly Notices of the Royal
  Astronomical Society, 461, 1590

\bibitem[{{Erkal} {et~al.}(2018){Erkal}, {Li}, {Koposov}, {Belokurov},
  {Balbinot}, {Bechtol}, {Buncher}, {Drlica-Wagner}, {Kuehn}, {Marshall},
  {Mart{\'\i}nez-V{\'a}zquez}, {Pace}, {Shipp}, {Simon}, {Stringer}, {Vivas},
  {Wechsler}, {Yanny}, {Abdalla}, {Allam}, {Annis}, {Avila}, {Bertin},
  {Brooks}, {Buckley-Geer}, {Burke}, {Carnero Rosell}, {Carrasco Kind},
  {Carretero}, {D'Andrea}, {da Costa}, {Davis}, {De Vicente}, {Doel}, {Eifler},
  {Evrard}, {Flaugher}, {Frieman}, {Garc{\'\i}a-Bellido}, {Gaztanaga},
  {Gerdes}, {Gruen}, {Gruendl}, {Gschwend}, {Gutierrez}, {Hartley},
  {Hollowood}, {Honscheid}, {James}, {Krause}, {Maia}, {March}, {Menanteau},
  {Miquel}, {Ogando}, {Plazas}, {Sanchez}, {Santiago}, {Scarpine}, {Schindler},
  {Sevilla-Noarbe}, {Smith}, {Smith}, {Soares-Santos}, {Sobreira}, {Suchyta},
  {Swanson}, {Tarle}, {Tucker}, \& {Walker}}]{Erkal_2018_TucIII}
{Erkal}, D., {Li}, T.~S., {Koposov}, S.~E., {et~al.} 2018, \mnras, 481, 3148

\bibitem[{{Erkal} {et~al.}(2019){Erkal}, {Belokurov}, {Laporte}, {Koposov},
  {Li}, {Grillmair}, {Kallivayalil}, {Price-Whelan}, {Evans}, {Hawkins},
  {Hendel}, {Mateu}, {Navarro}, {del Pino}, {Slater}, {Sohn}, \& {Orphan Aspen
  Treasury Collaboration}}]{Erkal2019_Orphan}
{Erkal}, D., {Belokurov}, V., {Laporte}, C.~F.~P., {et~al.} 2019, \mnras, 487,
  2685

\bibitem[{{Forbes} \& {Bridges}(2010)}]{Forbes_2010}
{Forbes}, D.~A., \& {Bridges}, T. 2010, \mnras, 404, 1203

\bibitem[{{Gaia Collaboration} {et~al.}(2018){Gaia Collaboration}, {Brown, A.
  G. A.}, {Vallenari, A.}, {Prusti, T.}, {de Bruijne, J. H. J.}, \& {et
  al.}}]{GaiaDR2_2018_Brown}
{Gaia Collaboration}, {Brown, A. G. A.}, {Vallenari, A.}, {et~al.} 2018, A\&A,
  doi:10.1051/0004-6361/201833051

\bibitem[{{Gaia Collaboration} {et~al.}(2020){Gaia Collaboration}, {Brown,
  Anthony G.A.}, {Vallenari, A.}, {Prusti, T.}, \& {de Bruijne, J.
  H.J.}}]{GaiaEDR3_Brown_2020}
{Gaia Collaboration}, {Brown, Anthony G.A.}, {Vallenari, A.}, {Prusti, T.}, \&
  {de Bruijne, J. H.J.} 2020, A\&A, doi:10.1051/0004-6361/202039657

\bibitem[{{Gaia Collaboration} {et~al.}(2016){Gaia Collaboration}, {Prusti},
  {de Bruijne}, {Brown}, {Vallenari}, {Babusiaux}, {Bailer-Jones}, {Bastian},
  {Biermann}, {Evans}, {Eyer}, {Jansen}, {Jordi}, {Klioner}, {Lammers},
  {Lindegren}, {Luri}, {Mignard}, {Milligan}, {Panem}, {Poinsignon},
  {Pourbaix}, {Randich}, {Sarri}, {Sartoretti}, {Siddiqui}, {Soubiran},
  {Valette}, {van Leeuwen}, {Walton}, {Aerts}, {Arenou}, {Cropper}, {Drimmel},
  {H{\o}g}, {Katz}, {Lattanzi}, {O'Mullane}, {Grebel}, {Holland}, {Huc},
  {Passot}, {Bramante}, {Cacciari}, {Casta{\~n}eda}, {Chaoul}, {Cheek}, {De
  Angeli}, {Fabricius}, {Guerra}, {Hern{\'a}ndez}, {Jean-Antoine-Piccolo},
  {Masana}, {Messineo}, {Mowlavi}, {Nienartowicz}, {Ord{\'o}{\~n}ez-Blanco},
  {Panuzzo}, {Portell}, {Richards}, {Riello}, {Seabroke}, {Tanga},
  {Th{\'e}venin}, {Torra}, {Els}, {Gracia-Abril}, {Comoretto},
  {Garcia-Reinaldos}, {Lock}, {Mercier}, {Altmann}, {Andrae}, {Astraatmadja},
  {Bellas-Velidis}, {Benson}, {Berthier}, {Blomme}, {Busso}, {Carry},
  {Cellino}, {Clementini}, {Cowell}, {Creevey}, {Cuypers}, {Davidson}, {De
  Ridder}, {de Torres}, {Delchambre}, {Dell'Oro}, {Ducourant}, {Fr{\'e}mat},
  {Garc{\'\i}a-Torres}, {Gosset}, {Halbwachs}, {Hambly}, {Harrison}, {Hauser},
  {Hestroffer}, {Hodgkin}, {Huckle}, {Hutton}, {Jasniewicz}, {Jordan},
  {Kontizas}, {Korn}, {Lanzafame}, {Manteiga}, {Moitinho}, {Muinonen},
  {Osinde}, {Pancino}, {Pauwels}, {Petit}, {Recio-Blanco}, {Robin}, {Sarro},
  {Siopis}, {Smith}, {Smith}, {Sozzetti}, {Thuillot}, {van Reeven}, {Viala},
  {Abbas}, {Abreu Aramburu}, {Accart}, {Aguado}, {Allan}, {Allasia},
  {Altavilla}, {{\'A}lvarez}, {Alves}, {Anderson}, {Andrei}, {Anglada Varela},
  {Antiche}, {Antoja}, {Ant{\'o}n}, {Arcay}, {Atzei}, {Ayache}, {Bach},
  {Baker}, {Balaguer-N{\'u}{\~n}ez}, {Barache}, {Barata}, {Barbier}, {Barblan},
  {Baroni}, {Barrado y Navascu{\'e}s}, {Barros}, {Barstow}, {Becciani},
  {Bellazzini}, {Bellei}, {Bello Garc{\'\i}a}, {Belokurov}, {Bendjoya},
  {Berihuete}, {Bianchi}, {Bienaym{\'e}}, {Billebaud}, {Blagorodnova},
  {Blanco-Cuaresma}, {Boch}, {Bombrun}, {Borrachero}, {Bouquillon}, {Bourda},
  {Bouy}, {Bragaglia}, {Breddels}, {Brouillet}, {Br{\"u}semeister},
  {Bucciarelli}, {Budnik}, {Burgess}, {Burgon}, {Burlacu}, {Busonero}, {Buzzi},
  {Caffau}, {Cambras}, {Campbell}, {Cancelliere}, {Cantat-Gaudin}, {Carlucci},
  {Carrasco}, {Castellani}, {Charlot}, {Charnas}, {Charvet}, {Chassat},
  {Chiavassa}, {Clotet}, {Cocozza}, {Collins}, {Collins}, {Costigan}, {Crifo},
  {Cross}, {Crosta}, {Crowley}, {Dafonte}, {Damerdji}, {Dapergolas}, {David},
  {David}, {De Cat}, {de Felice}, {de Laverny}, {De Luise}, {De March}, {de
  Martino}, {de Souza}, {Debosscher}, {del Pozo}, {Delbo}, {Delgado},
  {Delgado}, {di Marco}, {Di Matteo}, {Diakite}, {Distefano}, {Dolding}, {Dos
  Anjos}, {Drazinos}, {Dur{\'a}n}, {Dzigan}, {Ecale}, {Edvardsson}, {Enke},
  {Erdmann}, {Escolar}, {Espina}, {Evans}, {Eynard Bontemps}, {Fabre},
  {Fabrizio}, {Faigler}, {Falc{\~a}o}, {Farr{\`a}s Casas}, {Faye}, {Federici},
  {Fedorets}, {Fern{\'a}ndez-Hern{\'a}ndez}, {Fernique}, {Fienga}, {Figueras},
  {Filippi}, {Findeisen}, {Fonti}, {Fouesneau}, {Fraile}, {Fraser}, {Fuchs},
  {Furnell}, {Gai}, {Galleti}, {Galluccio}, {Garabato}, {Garc{\'\i}a-Sedano},
  {Gar{\'e}}, {Garofalo}, {Garralda}, {Gavras}, {Gerssen}, {Geyer}, {Gilmore},
  {Girona}, {Giuffrida}, {Gomes}, {Gonz{\'a}lez-Marcos},
  {Gonz{\'a}lez-N{\'u}{\~n}ez}, {Gonz{\'a}lez-Vidal}, {Granvik}, {Guerrier},
  {Guillout}, {Guiraud}, {G{\'u}rpide}, {Guti{\'e}rrez-S{\'a}nchez}, {Guy},
  {Haigron}, {Hatzidimitriou}, {Haywood}, {Heiter}, {Helmi}, {Hobbs},
  {Hofmann}, {Holl}, {Holland}, {Hunt}, {Hypki}, {Icardi}, {Irwin}, {Jevardat
  de Fombelle}, {Jofr{\'e}}, {Jonker}, {Jorissen}, {Julbe}, {Karampelas},
  {Kochoska}, {Kohley}, {Kolenberg}, {Kontizas}, {Koposov}, {Kordopatis},
  {Koubsky}, {Kowalczyk}, {Krone-Martins}, {Kudryashova}, {Kull}, {Bachchan},
  {Lacoste-Seris}, {Lanza}, {Lavigne}, {Le Poncin-Lafitte}, {Lebreton},
  {Lebzelter}, {Leccia}, {Leclerc}, {Lecoeur-Taibi}, {Lemaitre}, {Lenhardt},
  {Leroux}, {Liao}, {Licata}, {Lindstr{\o}m}, {Lister}, {Livanou}, {Lobel},
  {L{\"o}ffler}, {L{\'o}pez}, {Lopez-Lozano}, {Lorenz}, {Loureiro},
  {MacDonald}, {Magalh{\~a}es Fernandes}, {Managau}, {Mann}, {Mantelet},
  {Marchal}, {Marchant}, {Marconi}, {Marie}, {Marinoni}, {Marrese},
  {Marschalk{\'o}}, {Marshall}, {Mart{\'\i}n-Fleitas}, {Martino}, {Mary},
  {Matijevi{\v{c}}}, {Mazeh}, {McMillan}, {Messina}, {Mestre}, {Michalik},
  {Millar}, {Miranda}, {Molina}, {Molinaro}, {Molinaro}, {Moln{\'a}r},
  {Moniez}, {Montegriffo}, {Monteiro}, {Mor}, {Mora}, {Morbidelli}, {Morel},
  {Morgenthaler}, {Morley}, {Morris}, {Mulone}, {Muraveva}, {Musella},
  {Narbonne}, {Nelemans}, {Nicastro}, {Noval}, {Ord{\'e}novic},
  {Ordieres-Mer{\'e}}, {Osborne}, {Pagani}, {Pagano}, {Pailler}, {Palacin},
  {Palaversa}, {Parsons}, {Paulsen}, {Pecoraro}, {Pedrosa}, {Pentik{\"a}inen},
  {Pereira}, {Pichon}, {Piersimoni}, {Pineau}, {Plachy}, {Plum}, {Poujoulet},
  {Pr{\v{s}}a}, {Pulone}, {Ragaini}, {Rago}, {Rambaux}, {Ramos-Lerate},
  {Ranalli}, {Rauw}, {Read}, {Regibo}, {Renk}, {Reyl{\'e}}, {Ribeiro},
  {Rimoldini}, {Ripepi}, {Riva}, {Rixon}, {Roelens}, {Romero-G{\'o}mez},
  {Rowell}, {Royer}, {Rudolph}, {Ruiz-Dern}, {Sadowski}, {Sagrist{\`a}
  Sell{\'e}s}, {Sahlmann}, {Salgado}, {Salguero}, {Sarasso}, {Savietto},
  {Schnorhk}, {Schultheis}, {Sciacca}, {Segol}, {Segovia}, {Segransan},
  {Serpell}, {Shih}, {Smareglia}, {Smart}, {Smith}, {Solano}, {Solitro},
  {Sordo}, {Soria Nieto}, {Souchay}, {Spagna}, {Spoto}, {Stampa}, {Steele},
  {Steidelm{\"u}ller}, {Stephenson}, {Stoev}, {Suess}, {S{\"u}veges}, {Surdej},
  {Szabados}, {Szegedi-Elek}, {Tapiador}, {Taris}, {Tauran}, {Taylor},
  {Teixeira}, {Terrett}, {Tingley}, {Trager}, {Turon}, {Ulla}, {Utrilla},
  {Valentini}, {van Elteren}, {Van Hemelryck}, {van Leeuwen}, {Varadi},
  {Vecchiato}, {Veljanoski}, {Via}, {Vicente}, {Vogt}, {Voss}, {Votruba},
  {Voutsinas}, {Walmsley}, {Weiler}, {Weingrill}, {Werner}, {Wevers},
  {Whitehead}, {Wyrzykowski}, {Yoldas}, {{\v{Z}}erjal}, {Zucker}, {Zurbach},
  {Zwitter}, {Alecu}, {Allen}, {Allende Prieto}, {Amorim},
  {Anglada-Escud{\'e}}, {Arsenijevic}, {Azaz}, {Balm}, {Beck}, {Bernstein},
  {Bigot}, {Bijaoui}, {Blasco}, {Bonfigli}, {Bono}, {Boudreault}, {Bressan},
  {Brown}, {Brunet}, {Bunclark}, {Buonanno}, {Butkevich}, {Carret}, {Carrion},
  {Chemin}, {Ch{\'e}reau}, {Corcione}, {Darmigny}, {de Boer}, {de Teodoro}, {de
  Zeeuw}, {Delle Luche}, {Domingues}, {Dubath}, {Fodor}, {Fr{\'e}zouls},
  {Fries}, {Fustes}, {Fyfe}, {Gallardo}, {Gallegos}, {Gardiol}, {Gebran},
  {Gomboc}, {G{\'o}mez}, {Grux}, {Gueguen}, {Heyrovsky}, {Hoar}, {Iannicola},
  {Isasi Parache}, {Janotto}, {Joliet}, {Jonckheere}, {Keil}, {Kim},
  {Klagyivik}, {Klar}, {Knude}, {Kochukhov}, {Kolka}, {Kos}, {Kutka}, {Lainey},
  {LeBouquin}, {Liu}, {Loreggia}, {Makarov}, {Marseille}, {Martayan},
  {Martinez-Rubi}, {Massart}, {Meynadier}, {Mignot}, {Munari}, {Nguyen},
  {Nordlander}, {Ocvirk}, {O'Flaherty}, {Olias Sanz}, {Ortiz}, {Osorio},
  {Oszkiewicz}, {Ouzounis}, {Palmer}, {Park}, {Pasquato}, {Peltzer}, {Peralta},
  {P{\'e}turaud}, {Pieniluoma}, {Pigozzi}, {Poels}, {Prat}, {Prod'homme},
  {Raison}, {Rebordao}, {Risquez}, {Rocca-Volmerange}, {Rosen}, {Ruiz-Fuertes},
  {Russo}, {Sembay}, {Serraller Vizcaino}, {Short}, {Siebert}, {Silva},
  {Sinachopoulos}, {Slezak}, {Soffel}, {Sosnowska}, {Strai{\v{z}}ys}, {ter
  Linden}, {Terrell}, {Theil}, {Tiede}, {Troisi}, {Tsalmantza}, {Tur},
  {Vaccari}, {Vachier}, {Valles}, {Van Hamme}, {Veltz}, {Virtanen}, {Wallut},
  {Wichmann}, {Wilkinson}, {Ziaeepour}, \& {Zschocke}}]{Prusti_2016}
{Gaia Collaboration}, {Prusti}, T., {de Bruijne}, J.~H.~J., {et~al.} 2016,
  \aap, 595, A1

\bibitem[{{Gibbons} {et~al.}(2017){Gibbons}, {Belokurov}, \&
  {Evans}}]{Gibbons2017}
{Gibbons}, S.~L.~J., {Belokurov}, V., \& {Evans}, N.~W. 2017, \mnras, 464, 794

\bibitem[{{Gravity Collaboration} {et~al.}(2018){Gravity Collaboration},
  {Abuter}, {Amorim}, {Anugu}, {Baub{\"o}ck}, {Benisty}, {Berger}, {Blind},
  {Bonnet}, {Brandner}, {Buron}, {Collin}, {Chapron}, {Cl{\'e}net}, {Coud{\'e}
  Du Foresto}, {de Zeeuw}, {Deen}, {Delplancke-Str{\"o}bele}, {Dembet},
  {Dexter}, {Duvert}, {Eckart}, {Eisenhauer}, {Finger}, {F{\"o}rster
  Schreiber}, {F{\'e}dou}, {Garcia}, {Garcia Lopez}, {Gao}, {Gendron},
  {Genzel}, {Gillessen}, {Gordo}, {Habibi}, {Haubois}, {Haug}, {Hau{\ss}mann},
  {Henning}, {Hippler}, {Horrobin}, {Hubert}, {Hubin}, {Jimenez Rosales},
  {Jochum}, {Jocou}, {Kaufer}, {Kellner}, {Kendrew}, {Kervella}, {Kok},
  {Kulas}, {Lacour}, {Lapeyr{\`e}re}, {Lazareff}, {Le Bouquin}, {L{\'e}na},
  {Lippa}, {Lenzen}, {M{\'e}rand}, {M{\"u}ler}, {Neumann}, {Ott}, {Palanca},
  {Paumard}, {Pasquini}, {Perraut}, {Perrin}, {Pfuhl}, {Plewa}, {Rabien},
  {Ram{\'\i}rez}, {Ramos}, {Rau}, {Rodr{\'\i}guez-Coira}, {Rohloff}, {Rousset},
  {Sanchez-Bermudez}, {Scheithauer}, {Sch{\"o}ller}, {Schuler}, {Spyromilio},
  {Straub}, {Straubmeier}, {Sturm}, {Tacconi}, {Tristram}, {Vincent}, {von
  Fellenberg}, {Wank}, {Waisberg}, {Widmann}, {Wieprecht}, {Wiest},
  {Wiezorrek}, {Woillez}, {Yazici}, {Ziegler}, \& {Zins}}]{Gravity2018}
{Gravity Collaboration}, {Abuter}, R., {Amorim}, A., {et~al.} 2018, \aap, 615,
  L15

\bibitem[{{Grillmair}(2006)}]{Grillmair_Orphan2006}
{Grillmair}, C.~J. 2006, \apjl, 645, L37

\bibitem[{{Hayes} {et~al.}(2020){Hayes}, {Majewski}, {Hasselquist}, {Anguiano},
  {Shetrone}, {Law}, {Schiavon}, {Cunha}, {Smith}, {Beaton}, {Price-Whelan},
  {Allende Prieto}, {Battaglia}, {Bizyaev}, {Brownstein}, {Cohen},
  {Frinchaboy}, {Garc{\'\i}a-Hern{\'a}ndez}, {Lacerna}, {Lane},
  {M{\'e}sz{\'a}ros}, {Bidin}, {M{\~{u}}noz}, {Nidever}, {Oravetz}, {Oravetz},
  {Pan}, {Roman-Lopes}, {Sobeck}, \& {Stringfellow}}]{Hayes_2020_Sgr}
{Hayes}, C.~R., {Majewski}, S.~R., {Hasselquist}, S., {et~al.} 2020, \apj, 889,
  63

\bibitem[{{Helmi} {et~al.}(2018){Helmi}, {Babusiaux}, {Koppelman}, {Massari},
  {Veljanoski}, \& {Brown}}]{Helmi2018}
{Helmi}, A., {Babusiaux}, C., {Koppelman}, H.~H., {et~al.} 2018, \nat, 563, 85

\bibitem[{{Holtzman} {et~al.}(2018){Holtzman}, {Hasselquist}, {Shetrone},
  {Cunha}, {Allende Prieto}, {Anguiano}, {Bizyaev}, {Bovy}, {Casey},
  {Edvardsson}, {Johnson}, {J{\"o}nsson}, {Meszaros}, {Smith}, {Sobeck},
  {Zamora}, {Chojnowski}, {Fernandez-Trincado}, {Garcia-Hernandez}, {Majewski},
  {Pinsonneault}, {Souto}, {Stringfellow}, {Tayar}, {Troup}, \&
  {Zasowski}}]{holtzman18}
{Holtzman}, J.~A., {Hasselquist}, S., {Shetrone}, M., {et~al.} 2018, \aj, 156,
  125

\bibitem[{{Hudson} {et~al.}(2014){Hudson}, {Harris}, \& {Harris}}]{Hudson2014}
{Hudson}, M.~J., {Harris}, G.~L., \& {Harris}, W.~E. 2014, \apjl, 787, L5

\bibitem[{{Ibata} {et~al.}(2020{\natexlab{a}}){Ibata}, {Bellazzini}, {Thomas},
  {Malhan}, {Martin}, {Famaey}, \& {Siebert}}]{Ibata_Sgr2020}
{Ibata}, R., {Bellazzini}, M., {Thomas}, G., {et~al.} 2020{\natexlab{a}},
  \apjl, 891, L19

\bibitem[{{Ibata} {et~al.}(2020{\natexlab{b}}){Ibata}, {Bellazzini}, {Thomas},
  {Malhan}, {Martin}, {Famaey}, \& {Siebert}}]{Ibata_Sgr_2020}
---. 2020{\natexlab{b}}, \apjl, 891, L19

\bibitem[{{Ibata} {et~al.}(2020{\natexlab{c}}){Ibata}, {Malhan}, {Martin},
  {Aubert}, {Famaey}, {Bianchini}, {Monari}, {Siebert}, {Thomas}, {Bellazzini},
  {Bonifacio}, {Caffau}, \& {Renaud}}]{Ibata_GaiaEDR3}
{Ibata}, R., {Malhan}, K., {Martin}, N., {et~al.} 2020{\natexlab{c}}, arXiv
  e-prints, arXiv:2012.05245

\bibitem[{{Ibata} {et~al.}(2003){Ibata}, {Irwin}, {Lewis}, {Ferguson}, \&
  {Tanvir}}]{Ibata_GASS_2003}
{Ibata}, R.~A., {Irwin}, M.~J., {Lewis}, G.~F., {Ferguson}, A.~M.~N., \&
  {Tanvir}, N. 2003, \mnras, 340, L21

\bibitem[{{Ibata} {et~al.}(2019){Ibata}, {Malhan}, \&
  {Martin}}]{Ibata_norse_2019}
{Ibata}, R.~A., {Malhan}, K., \& {Martin}, N.~F. 2019, \apj, 872, 152

\bibitem[{{Ji} {et~al.}(2020){Ji}, {Li}, {Hansen}, {Casey}, {Koposov}, {Pace},
  {Mackey}, {Lewis}, {Simpson}, {Bland-Hawthorn}, {Cullinane}, {Da Costa},
  {Hattori}, {Martell}, {Kuehn}, {Erkal}, {Shipp}, {Wan}, \&
  {Zucker}}]{Ji_S5_2020}
{Ji}, A.~P., {Li}, T.~S., {Hansen}, T.~T., {et~al.} 2020, \aj, 160, 181

\bibitem[{{Jofr{\'e}} {et~al.}(2019){Jofr{\'e}}, {Heiter}, \&
  {Soubiran}}]{Jofre2019}
{Jofr{\'e}}, P., {Heiter}, U., \& {Soubiran}, C. 2019, \araa, 57, 571

\bibitem[{{Kirby} {et~al.}(2013){Kirby}, {Cohen}, {Guhathakurta}, {Cheng},
  {Bullock}, \& {Gallazzi}}]{Kirby2013}
{Kirby}, E.~N., {Cohen}, J.~G., {Guhathakurta}, P., {et~al.} 2013, \apj, 779,
  102

\bibitem[{{Kirby} {et~al.}(2011){Kirby}, {Lanfranchi}, {Simon}, {Cohen}, \&
  {Guhathakurta}}]{Kirby2011}
{Kirby}, E.~N., {Lanfranchi}, G.~A., {Simon}, J.~D., {Cohen}, J.~G., \&
  {Guhathakurta}, P. 2011, \apj, 727, 78

\bibitem[{{Koch} \& {C{\^o}t{\'e}}(2017)}]{Koch2017}
{Koch}, A., \& {C{\^o}t{\'e}}, P. 2017, \aap, 601, A41

\bibitem[{{Koleva} {et~al.}(2009){Koleva}, {Prugniel}, {Bouchard}, \&
  {Wu}}]{Koleva09}
{Koleva}, M., {Prugniel}, P., {Bouchard}, A., \& {Wu}, Y. 2009, \aap, 501, 1269

\bibitem[{{Koposov} {et~al.}(2019){Koposov}, {Belokurov}, {Li}, {Mateu},
  {Erkal}, {Grillmair}, {Hendel}, {Price-Whelan}, {Laporte}, {Hawkins}, {Sohn},
  {del Pino}, {Evans}, {Slater}, {Kallivayalil}, {Navarro}, \& {Orphan Aspen
  Treasury Collaboration}}]{Koposov2019}
{Koposov}, S.~E., {Belokurov}, V., {Li}, T.~S., {et~al.} 2019, \mnras, 485,
  4726

\bibitem[{{Koppelman} {et~al.}(2019){Koppelman}, {Helmi}, {Massari},
  {Price-Whelan}, \& {Starkenburg}}]{Koppelman2019}
{Koppelman}, H.~H., {Helmi}, A., {Massari}, D., {Price-Whelan}, A.~M., \&
  {Starkenburg}, T.~K. 2019, \aap, 631, L9

\bibitem[{{Kruijssen} {et~al.}(2019){Kruijssen}, {Pfeffer}, {Reina-Campos},
  {Crain}, \& {Bastian}}]{Kruijssen_2019}
{Kruijssen}, J.~M.~D., {Pfeffer}, J.~L., {Reina-Campos}, M., {Crain}, R.~A., \&
  {Bastian}, N. 2019, \mnras, 486, 3180

\bibitem[{{Law} \& {Majewski}(2010)}]{Law2010}
{Law}, D.~R., \& {Majewski}, S.~R. 2010, \apj, 714, 229

\bibitem[{{Lee} {et~al.}(2008){Lee}, {Beers}, {Sivarani}, {Johnson}, {An},
  {Wilhelm}, {Allende Prieto}, {Koesterke}, {Re Fiorentin}, {Bailer-Jones},
  {Norris}, {Yanny}, {Rockosi}, {Newberg}, {Cudworth}, \& {Pan}}]{lee08}
{Lee}, Y.~S., {Beers}, T.~C., {Sivarani}, T., {et~al.} 2008, \aj, 136, 2050

\bibitem[{{Lee} {et~al.}(2015){Lee}, {Beers}, {Carlin}, {Newberg}, {Hou}, {Li},
  {Luo}, {Wu}, {Yang}, {Zhang}, {Zhang}, \& {Zhang}}]{lee15}
{Lee}, Y.~S., {Beers}, T.~C., {Carlin}, J.~L., {et~al.} 2015, \aj, 150, 187

\bibitem[{{Li} {et~al.}(2021){Li}, {Hammer}, {Babusiaux}, {Pawlowski}, {Yang},
  {Arenou}, {Du}, \& {Wang}}]{Li_MWdwarfs_GaiaEDR3}
{Li}, H., {Hammer}, F., {Babusiaux}, C., {et~al.} 2021, arXiv e-prints,
  arXiv:2104.03974

\bibitem[{{Li} {et~al.}(2018{\natexlab{a}}){Li}, {Tan}, \&
  {Zhao}}]{Li_LAMOSTDR32018}
{Li}, H., {Tan}, K., \& {Zhao}, G. 2018{\natexlab{a}}, \apjs, 238, 16

\bibitem[{{Li} {et~al.}(2018{\natexlab{b}}){Li}, {Simon}, {Kuehn}, {Pace},
  {Erkal}, {Bechtol}, {Yanny}, {Drlica-Wagner}, {Marshall}, {Lidman},
  {Balbinot}, {Carollo}, {Jenkins}, {Mart{\'\i}nez-V{\'a}zquez}, {Shipp},
  {Stringer}, {Vivas}, {Walker}, {Wechsler}, {Abdalla}, {Allam}, {Annis},
  {Avila}, {Bertin}, {Brooks}, {Buckley-Geer}, {Burke}, {Carnero Rosell},
  {Carrasco Kind}, {Carretero}, {Cunha}, {D'Andrea}, {da Costa}, {Davis}, {De
  Vicente}, {Doel}, {Eifler}, {Evrard}, {Flaugher}, {Frieman},
  {Garc{\'\i}a-Bellido}, {Gaztanaga}, {Gerdes}, {Gruen}, {Gruendl}, {Gschwend},
  {Gutierrez}, {Hartley}, {Hollowood}, {Honscheid}, {James}, {Krause}, {Maia},
  {March}, {Menanteau}, {Miquel}, {Plazas}, {Sanchez}, {Santiago}, {Scarpine},
  {Schindler}, {Schubnell}, {Sevilla-Noarbe}, {Smith}, {Smith},
  {Soares-Santos}, {Sobreira}, {Suchyta}, {Swanson}, {Tarle}, {Tucker}, \& {DES
  Collaboration}}]{Li_TucIII_stream2018}
{Li}, T.~S., {Simon}, J.~D., {Kuehn}, K., {et~al.} 2018{\natexlab{b}}, \apj,
  866, 22

\bibitem[{{Lian} {et~al.}(2020){Lian}, {Thomas}, {Maraston}, {Beers}, {Moni
  Bidin}, {Fern{\'a}ndez-Trincado}, {Garc{\'\i}a-Hern{\'a}ndez}, {Lane},
  {Munoz}, {Nitschelm}, {Roman-Lopes}, \& {Zamora}}]{Lian2020}
{Lian}, J., {Thomas}, D., {Maraston}, C., {et~al.} 2020, \mnras, 497, 2371

\bibitem[{{Lindegren} {et~al.}(2018){Lindegren}, {Hern{\'a}ndez}, {Bombrun},
  {Klioner}, {Bastian}, {Ramos-Lerate}, {de Torres}, \&
  {Steidelm{\"u}ller}}]{Lindegren2018}
{Lindegren}, L., {Hern{\'a}ndez}, J., {Bombrun}, A., {et~al.} 2018, \aap, 616,
  A2

\bibitem[{{Majewski} {et~al.}(2003){Majewski}, {Skrutskie}, {Weinberg}, \&
  {Ostheimer}}]{Majewski2003}
{Majewski}, S.~R., {Skrutskie}, M.~F., {Weinberg}, M.~D., \& {Ostheimer}, J.~C.
  2003, \apj, 599, 1082

\bibitem[{{Majewski} {et~al.}(2017){Majewski}, {Schiavon}, {Frinchaboy},
  {Allende Prieto}, {Barkhouser}, {Bizyaev}, {Blank}, {Brunner}, {Burton},
  {Carrera}, {Chojnowski}, {Cunha}, {Epstein}, {Fitzgerald}, {Garc{\'\i}a
  P{\'e}rez}, {Hearty}, {Henderson}, {Holtzman}, {Johnson}, {Lam}, {Lawler},
  {Maseman}, {M{\'e}sz{\'a}ros}, {Nelson}, {Nguyen}, {Nidever}, {Pinsonneault},
  {Shetrone}, {Smee}, {Smith}, {Stolberg}, {Skrutskie}, {Walker}, {Wilson},
  {Zasowski}, {Anders}, {Basu}, {Beland}, {Blanton}, {Bovy}, {Brownstein},
  {Carlberg}, {Chaplin}, {Chiappini}, {Eisenstein}, {Elsworth}, {Feuillet},
  {Fleming}, {Galbraith-Frew}, {Garc{\'\i}a}, {Garc{\'\i}a-Hern{\'a}ndez},
  {Gillespie}, {Girardi}, {Gunn}, {Hasselquist}, {Hayden}, {Hekker}, {Ivans},
  {Kinemuchi}, {Klaene}, {Mahadevan}, {Mathur}, {Mosser}, {Muna}, {Munn},
  {Nichol}, {O'Connell}, {Parejko}, {Robin}, {Rocha-Pinto}, {Schultheis},
  {Serenelli}, {Shane}, {Silva Aguirre}, {Sobeck}, {Thompson}, {Troup},
  {Weinberg}, \& {Zamora}}]{Majewski_APOGEE_2017}
{Majewski}, S.~R., {Schiavon}, R.~P., {Frinchaboy}, P.~M., {et~al.} 2017, \aj,
  154, 94

\bibitem[{{Malhan} \& {Ibata}(2018)}]{Malhan_SF_2018}
{Malhan}, K., \& {Ibata}, R.~A. 2018, \mnras, 477, 4063

\bibitem[{{Malhan} \& {Ibata}(2019)}]{Malhan_GalPot_2019}
---. 2019, \mnras, 486, 2995

\bibitem[{{Malhan} {et~al.}(2019){Malhan}, {Ibata}, {Carlberg}, {Bellazzini},
  {Famaey}, \& {Martin}}]{Malhan_Kshir2019}
{Malhan}, K., {Ibata}, R.~A., {Carlberg}, R.~G., {et~al.} 2019, \apjl, 886, L7

\bibitem[{{Malhan} {et~al.}(2018{\natexlab{a}}){Malhan}, {Ibata}, {Goldman},
  {Martin}, {Magnier}, \& {Chambers}}]{Malhan_PS1_2018}
{Malhan}, K., {Ibata}, R.~A., {Goldman}, B., {et~al.} 2018{\natexlab{a}},
  \mnras, 478, 3862

\bibitem[{{Malhan} {et~al.}(2018{\natexlab{b}}){Malhan}, {Ibata}, \&
  {Martin}}]{Malhan_Ghostly_2018}
{Malhan}, K., {Ibata}, R.~A., \& {Martin}, N.~F. 2018{\natexlab{b}}, \mnras,
  481, 3442

\bibitem[{{Massari} {et~al.}(2019){Massari}, {Koppelman}, \&
  {Helmi}}]{Massari2019}
{Massari}, D., {Koppelman}, H.~H., \& {Helmi}, A. 2019, \aap, 630, L4

\bibitem[{{Matsuno} {et~al.}(2019){Matsuno}, {Aoki}, \& {Suda}}]{Matsuno2019}
{Matsuno}, T., {Aoki}, W., \& {Suda}, T. 2019, \apjl, 874, L35

\bibitem[{{McConnachie}(2012)}]{McConnachie2012}
{McConnachie}, A.~W. 2012, \aj, 144, 4

\bibitem[{{McMillan}(2017)}]{McMillan2017}
{McMillan}, P.~J. 2017, \mnras, 465, 76

\bibitem[{{Milone} {et~al.}(2017){Milone}, {Piotto}, {Renzini}, {Marino},
  {Bedin}, {Vesperini}, {D'Antona}, {Nardiello}, {Anderson}, {King}, {Yong},
  {Bellini}, {Aparicio}, {Barbuy}, {Brown}, {Cassisi}, {Ortolani}, {Salaris},
  {Sarajedini}, \& {van der Marel}}]{Milone2017}
{Milone}, A.~P., {Piotto}, G., {Renzini}, A., {et~al.} 2017, \mnras, 464, 3636

\bibitem[{{Myeong} {et~al.}(2019){Myeong}, {Vasiliev}, {Iorio}, {Evans}, \&
  {Belokurov}}]{Myeong2019}
{Myeong}, G.~C., {Vasiliev}, E., {Iorio}, G., {Evans}, N.~W., \& {Belokurov},
  V. 2019, \mnras, 488, 1235

\bibitem[{{Naidu} {et~al.}(2020){Naidu}, {Conroy}, {Bonaca}, {Johnson}, {Ting},
  {Caldwell}, {Zaritsky}, \& {Cargile}}]{Naidu2020}
{Naidu}, R.~P., {Conroy}, C., {Bonaca}, A., {et~al.} 2020, \apj, 901, 48

\bibitem[{{Newberg} {et~al.}(2010){Newberg}, {Willett}, {Yanny}, \&
  {Xu}}]{Newberg_orphan_2010}
{Newberg}, H.~J., {Willett}, B.~A., {Yanny}, B., \& {Xu}, Y. 2010, \apj, 711,
  32

\bibitem[{{Newberg} {et~al.}(2009){Newberg}, {Yanny}, \&
  {Willett}}]{Newberg_Cetus2009}
{Newberg}, H.~J., {Yanny}, B., \& {Willett}, B.~A. 2009, \apjl, 700, L61

\bibitem[{{Newberg} {et~al.}(2002{\natexlab{a}}){Newberg}, {Yanny}, {Rockosi},
  {Grebel}, {Rix}, {Brinkmann}, {Csabai}, {Hennessy}, {Hindsley}, {Ibata},
  {Ivezi{\'c}}, {Lamb}, {Nash}, {Odenkirchen}, {Rave}, {Schneider}, {Smith},
  {Stolte}, \& {York}}]{Newberg_GASS_2002}
{Newberg}, H.~J., {Yanny}, B., {Rockosi}, C., {et~al.} 2002{\natexlab{a}},
  \apj, 569, 245

\bibitem[{{Newberg} {et~al.}(2002{\natexlab{b}}){Newberg}, {Yanny}, {Rockosi},
  {Grebel}, {Rix}, {Brinkmann}, {Csabai}, {Hennessy}, {Hindsley}, {Ibata},
  {Ivezi{\'c}}, {Lamb}, {Nash}, {Odenkirchen}, {Rave}, {Schneider}, {Smith},
  {Stolte}, \& {York}}]{Newberg2002}
---. 2002{\natexlab{b}}, \apj, 569, 245

\bibitem[{{Nissen} \& {Schuster}(2010)}]{Nissen2010}
{Nissen}, P.~E., \& {Schuster}, W.~J. 2010, \aap, 511, L10

\bibitem[{{Pfeffer} {et~al.}(2021){Pfeffer}, {Lardo}, {Bastian}, {Saracino}, \&
  {Kamann}}]{Pfeffer2021}
{Pfeffer}, J., {Lardo}, C., {Bastian}, N., {Saracino}, S., \& {Kamann}, S.
  2021, \mnras, 500, 2514

\bibitem[{{Pillepich} {et~al.}(2018){Pillepich}, {Springel}, {Nelson}, {Genel},
  {Naiman}, {Pakmor}, {Hernquist}, {Torrey}, {Vogelsberger}, {Weinberger}, \&
  {Marinacci}}]{Pillepich2018}
{Pillepich}, A., {Springel}, V., {Nelson}, D., {et~al.} 2018, \mnras, 473, 4077

\bibitem[{{Price-Whelan}(2017)}]{Price-Whelan_GALA_2017}
{Price-Whelan}, A.~M. 2017, The Journal of Open Source Software, 2, 388

\bibitem[{{Prugniel} {et~al.}(2011){Prugniel}, {Vauglin}, \&
  {Koleva}}]{prugniel11}
{Prugniel}, P., {Vauglin}, I., \& {Koleva}, M. 2011, \aap, 531, A165

\bibitem[{{Ramos} {et~al.}(2020){Ramos}, {Mateu}, {Antoja}, {Helmi},
  {Castro-Ginard}, {Balbinot}, \& {Carrasco}}]{Ramos2020}
{Ramos}, P., {Mateu}, C., {Antoja}, T., {et~al.} 2020, \aap, 638, A104

\bibitem[{{Read} {et~al.}(2017){Read}, {Iorio}, {Agertz}, \&
  {Fraternali}}]{Read2017}
{Read}, J.~I., {Iorio}, G., {Agertz}, O., \& {Fraternali}, F. 2017, \mnras,
  467, 2019

\bibitem[{{Reid} {et~al.}(2014){Reid}, {Menten}, {Brunthaler}, {Zheng}, {Dame},
  {Xu}, {Wu}, {Zhang}, {Sanna}, {Sato}, {Hachisuka}, {Choi}, {Immer},
  {Moscadelli}, {Rygl}, \& {Bartkiewicz}}]{Reid2014_Sun}
{Reid}, M.~J., {Menten}, K.~M., {Brunthaler}, A., {et~al.} 2014, \apj, 783, 130

\bibitem[{{Ruiz-Lara} {et~al.}(2020){Ruiz-Lara}, {Gallart}, {Bernard}, \&
  {Cassisi}}]{Ruiz-Lara2020}
{Ruiz-Lara}, T., {Gallart}, C., {Bernard}, E.~J., \& {Cassisi}, S. 2020, Nature
  Astronomy, 4, 965

\bibitem[{{Schlafly} \& {Finkbeiner}(2011)}]{Schlafly2011}
{Schlafly}, E.~F., \& {Finkbeiner}, D.~P. 2011, \apj, 737, 103

\bibitem[{{Schlegel} {et~al.}(1998){Schlegel}, {Finkbeiner}, \&
  {Davis}}]{Schlegel1998}
{Schlegel}, D.~J., {Finkbeiner}, D.~P., \& {Davis}, M. 1998, \apj, 500, 525

\bibitem[{{Sch{\"o}nrich} {et~al.}(2010){Sch{\"o}nrich}, {Binney}, \&
  {Dehnen}}]{Schornich2010_Sun}
{Sch{\"o}nrich}, R., {Binney}, J., \& {Dehnen}, W. 2010, \mnras, 403, 1829

\bibitem[{{Sharma} {et~al.}(2016){Sharma}, {Prugniel}, \& {Singh}}]{sharma16}
{Sharma}, K., {Prugniel}, P., \& {Singh}, H.~P. 2016, \aap, 585, A64

\bibitem[{{Shipp} {et~al.}(2018){Shipp}, {Drlica-Wagner}, {Balbinot},
  {Ferguson}, {Erkal}, {Li}, {Bechtol}, {Belokurov}, {Buncher}, {Carollo},
  {Carrasco Kind}, {Kuehn}, {Marshall}, {Pace}, {Rykoff}, {Sevilla-Noarbe},
  {Sheldon}, {Strigari}, {Vivas}, {Yanny}, {Zenteno}, {Abbott}, {Abdalla},
  {Allam}, {Avila}, {Bertin}, {Brooks}, {Burke}, {Carretero}, {Castander},
  {Cawthon}, {Crocce}, {Cunha}, {D'Andrea}, {da Costa}, {Davis}, {De Vicente},
  {Desai}, {Diehl}, {Doel}, {Evrard}, {Flaugher}, {Fosalba}, {Frieman},
  {Garc{\'\i}a-Bellido}, {Gaztanaga}, {Gerdes}, {Gruen}, {Gruendl}, {Gschwend},
  {Gutierrez}, {Hartley}, {Honscheid}, {Hoyle}, {James}, {Johnson}, {Krause},
  {Kuropatkin}, {Lahav}, {Lin}, {Maia}, {March}, {Martini}, {Menanteau},
  {Miller}, {Miquel}, {Nichol}, {Plazas}, {Romer}, {Sako}, {Sanchez},
  {Santiago}, {Scarpine}, {Schindler}, {Schubnell}, {Smith}, {Smith},
  {Sobreira}, {Suchyta}, {Swanson}, {Tarle}, {Thomas}, {Tucker}, {Walker},
  {Wechsler}, \& {DES Collaboration}}]{Shipp2018}
{Shipp}, N., {Drlica-Wagner}, A., {Balbinot}, E., {et~al.} 2018, \apj, 862, 114

\bibitem[{Sivia(1996)}]{sivia1996data}
Sivia, D. 1996, Data Analysis: A Bayesian Tutorial, Oxford science publications
  (Clarendon Press)

\bibitem[{{Slater} {et~al.}(2014){Slater}, {Bell}, {Schlafly}, {Morganson},
  {Martin}, {Rix}, {Pe{\~n}arrubia}, {Bernard}, {Ferguson}, {Martinez-Delgado},
  {Wyse}, {Burgett}, {Chambers}, {Draper}, {Hodapp}, {Kaiser}, {Magnier},
  {Metcalfe}, {Price}, {Tonry}, {Wainscoat}, \& {Waters}}]{Slater2014}
{Slater}, C.~T., {Bell}, E.~F., {Schlafly}, E.~F., {et~al.} 2014, \apj, 791, 9

\bibitem[{Torrey {et~al.}(2019)Torrey, Vogelsberger, Marinacci, Pakmor,
  Springel, Nelson, Naiman, Pillepich, Genel, Weinberger, \&
  Hernquist}]{Torrey_2019}
Torrey, P., Vogelsberger, M., Marinacci, F., {et~al.} 2019, Monthly Notices of
  the Royal Astronomical Society, 484, 5587

\bibitem[{{Xiang} {et~al.}(2019){Xiang}, {Ting}, {Rix}, {Sandford}, {Buder},
  {Lind}, {Liu}, {Shi}, \& {Zhang}}]{Xiang_LAMOSTDR5_2019}
{Xiang}, M., {Ting}, Y.-S., {Rix}, H.-W., {et~al.} 2019, \apjs, 245, 34

\bibitem[{Yam {et~al.}(2013)Yam, Carlin, Newberg, Dumas, OMalley, Newby, \&
  Martin}]{Yam_2013}
Yam, W., Carlin, J.~L., Newberg, H.~J., {et~al.} 2013, The Astrophysical
  Journal, 776, 133

\bibitem[{{Yanny} {et~al.}(2009){Yanny}, {Rockosi}, {Newberg}, {Knapp},
  {Adelman-McCarthy}, {Alcorn}, {Allam}, {Allende Prieto}, {An}, {Anderson},
  {Anderson}, {Bailer-Jones}, {Bastian}, {Beers}, {Bell}, {Belokurov},
  {Bizyaev}, {Blythe}, {Bochanski}, {Boroski}, {Brinchmann}, {Brinkmann},
  {Brewington}, {Carey}, {Cudworth}, {Evans}, {Evans}, {Gates}, {G{\"a}nsicke},
  {Gillespie}, {Gilmore}, {Nebot Gomez-Moran}, {Grebel}, {Greenwell}, {Gunn},
  {Jordan}, {Jordan}, {Harding}, {Harris}, {Hendry}, {Holder}, {Ivans},
  {Ivezi{\v{c}}}, {Jester}, {Johnson}, {Kent}, {Kleinman}, {Kniazev},
  {Krzesinski}, {Kron}, {Kuropatkin}, {Lebedeva}, {Lee}, {French Leger},
  {L{\'e}pine}, {Levine}, {Lin}, {Long}, {Loomis}, {Lupton}, {Malanushenko},
  {Malanushenko}, {Margon}, {Martinez-Delgado}, {McGehee}, {Monet}, {Morrison},
  {Munn}, {Neilsen}, {Nitta}, {Norris}, {Oravetz}, {Owen}, {Padmanabhan},
  {Pan}, {Peterson}, {Pier}, {Platson}, {Re Fiorentin}, {Richards}, {Rix},
  {Schlegel}, {Schneider}, {Schreiber}, {Schwope}, {Sibley}, {Simmons},
  {Snedden}, {Allyn Smith}, {Stark}, {Stauffer}, {Steinmetz}, {Stoughton},
  {SubbaRao}, {Szalay}, {Szkody}, {Thakar}, {Sivarani}, {Tucker}, {Uomoto},
  {Vanden Berk}, {Vidrih}, {Wadadekar}, {Watters}, {Wilhelm}, {Wyse}, {Yarger},
  \& {Zucker}}]{Yanny2009}
{Yanny}, B., {Rockosi}, C., {Newberg}, H.~J., {et~al.} 2009, \aj, 137, 4377

\bibitem[{{Yuan} {et~al.}(2020{\natexlab{a}}){Yuan}, {Chang}, {Beers}, \&
  {Huang}}]{Yuan2020}
{Yuan}, Z., {Chang}, J., {Beers}, T.~C., \& {Huang}, Y. 2020{\natexlab{a}},
  \apjl, 898, L37

\bibitem[{{Yuan} {et~al.}(2019){Yuan}, {Smith}, {Xue}, {Li}, {Liu}, {Wang},
  {Li}, \& {Chang}}]{Yuan2019_CPS}
{Yuan}, Z., {Smith}, M.~C., {Xue}, X.-X., {et~al.} 2019, \apj, 881, 164

\bibitem[{{Yuan} {et~al.}(2020{\natexlab{b}}){Yuan}, {Myeong}, {Beers},
  {Evans}, {Lee}, {Banerjee}, {Gudin}, {Hattori}, {Li}, {Matsuno}, {Placco},
  {Smith}, {Whitten}, \& {Zhao}}]{Yuan2020a}
{Yuan}, Z., {Myeong}, G.~C., {Beers}, T.~C., {et~al.} 2020{\natexlab{b}}, \apj,
  891, 39

\bibitem[{{Zhao} {et~al.}(2009){Zhao}, {Jing}, {Mo}, \&
  {B{\"o}rner}}]{Zhao_2009}
{Zhao}, D.~H., {Jing}, Y.~P., {Mo}, H.~J., \& {B{\"o}rner}, G. 2009, \apj, 707,
  354

\bibitem[{{Zhao} {et~al.}(2012){Zhao}, {Zhao}, {Chu}, {Jing}, \&
  {Deng}}]{Zhao2012}
{Zhao}, G., {Zhao}, Y.-H., {Chu}, Y.-Q., {Jing}, Y.-P., \& {Deng}, L.-C. 2012,
  Research in Astronomy and Astrophysics, 12, 723

\end{thebibliography}

\appendix

\section{Comparing [Fe/H] measurements for very metal-poor stars from different datasets}\label{appendix:FeH_comparison}

\begin{figure*}
\begin{center}
\includegraphics[width=0.88\hsize]{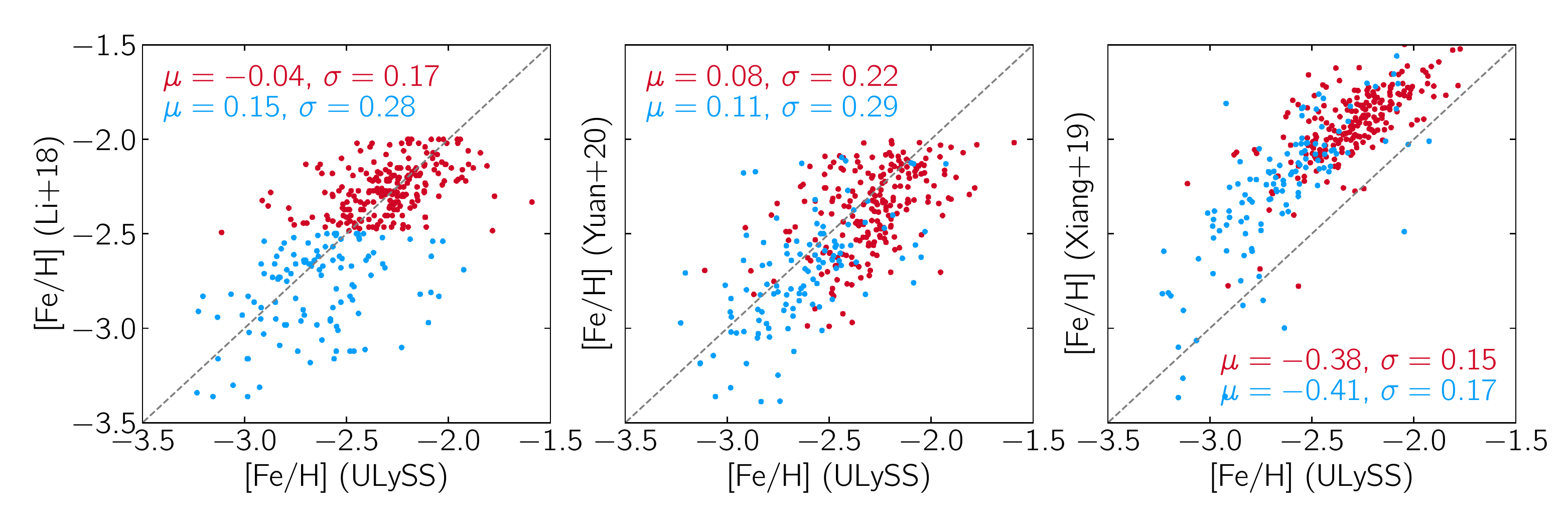}
\end{center}
\vspace{-0.7cm}
\caption{Comparing [Fe/H] from our method (ULySS) with [Fe/H] from different literature sources for a sample of very metal-poor LAMOST giants from \cite{Li_LAMOSTDR32018}. The literature values come from \cite{Li_LAMOSTDR32018}, \cite{Yuan2020a} and \cite{Xiang_LAMOSTDR5_2019} shown in the left-hand, middle and right-hand panel, respectively. The one-to-one relation is represented by the grey-dashed line, and the text in each panel presents the mean difference ($\mu$, ULySS minus literature) and dispersion ($\sigma$) between the samples, for stars with [Fe/H]$_\mathrm{Li+18} > -2.5$ (red) and $<-2.5$ (blue).}
\label{fig:compare-feh}
\end{figure*}

\subsection{Testing our spectroscopic metallicities}

The metallicities for LAMOST stars in this work were determined with the ULySS \citep{Koleva09} pipeline, making use of the empirical MILES library interpolator model \citep{prugniel11, sharma16}. This method has been tested extensively for stars of widely varying stellar types and metallicities, and was also used on the X-shooter Spectral Library \citep{Arentsen_2019}. Here we show that this method works well for estimating [Fe/H] of very metal-poor (VMP, $-3.0<$ [Fe/H] $<-2.0$) giant stars. We analyse a few hundred stars from the LAMOST VMP catalogue of \cite{Li_LAMOSTDR32018} in the same way as we analysed the LMS-1 stars, and present the comparison in the left-most panel of Figure~\ref{fig:compare-feh}. The agreement is excellent for $-2.5<$[Fe/H]$<-2.0$ with an average difference of $\mu \sim -0.04$~dex and a standard deviation of $\sigma \sim 0.17$~dex. For [Fe/H]$< -2.5$, the ULySS metallicities are on average $0.15$~dex higher, and there is a larger scatter of $0.28$~dex. The middle panel of Figure~\ref{fig:compare-feh} shows a comparison between our [Fe/H] estimates and those from \cite{Yuan2020a} who re-analysed the LAMOST spectra for the \cite{Li_LAMOSTDR32018} VMP sample with the Segue Stellar Parameter Pipeline for LAMOST (L-SSPP, \citealt{lee15}). There is a small offset that is similar in both metallicity ranges ($0.08/0.11$~dex), and the scatter is of the same order as before.

We also compare our [Fe/H] with those from \cite{Xiang_LAMOSTDR5_2019}, who analysed the LAMOST DR5 spectra using the ``data-driven Payne'' method, that they train on the parameters from APOGEE DR14 \citep{holtzman18} and GALAH DR2 \citep{buder18}. However, since the latter two surveys do not contain many metal-poor stars, the authors of \cite{Xiang_LAMOSTDR5_2019} themselves caution about their results for stars with [Fe/H] $<-1.5$. We confirm that this caution is necessary, as we find a large difference in metallicity ($\mu \sim -0.40$~dex) between our results and those from \cite{Xiang_LAMOSTDR5_2019} for a sample of VMP stars from the \cite{Li_LAMOSTDR32018} catalogue (see the right-most panel of Figure~\ref{fig:compare-feh}). We note that the dispersion is lower than in the previous two comparisons, even for [Fe/H]$< -2.5$. The reason for this low dispersion may owe to the fact that both the ULySS and the \cite{Xiang_LAMOSTDR5_2019} analysis are data-driven, and such techniques are known to provide more precise results \citep{Jofre2019}, although not necessarily more accurate. In sum, the overall comparison of our ULySS metallicities with \cite{Li_LAMOSTDR32018} and \cite{Yuan2020a} makes a strong case that our [Fe/H] measurements for the metal-poor stars are more accurate than those from \cite{Xiang_LAMOSTDR5_2019}. 

\subsection{Investigating differences in the metallicity scale of our method and the H3 survey}

LMS-1 and Wukong appear to be the same structure (especially in the dynamical space), but we find that our [Fe/H] measurement for LMS-1 is $\sim 0.4$ dex lower as compared to that of \cite{Naidu2020} for Wukong. This discrepancy may partly owe to the differences in the metallicity scales of the two analysis, and this discrepancy can be particularly strong for VMP stars. \cite{Naidu2020} derive the  median metallicity for Wukong using the H3 spectroscopic survey. \cite{cargile20} test the H3 pipeline on metal-poor globular clusters ([Fe/H] $\leq -1.5$) and find slightly higher metallicities than the literature (by $\sim 0.1-0.15$ dex), although they do find a good agreement with the literature specifically for the metal-poor ``Sextans'' dwarf galaxy (that has [Fe/H] $= -1.9$). We also note that small scale differences of $\sim 0.1-0.2$~dex generally exist across different surveys, and is a common issue since different spectroscopic surveys analyse different resolutions/wavelengths to derive chemical composition of stars. Additionally, \cite{Conroy_H3_2019} compare the metallicities from H3 with the literature, which for metal-poor stars comes from SEGUE DR14 (analysed with the SSPP, \citealt{lee08}) and from the \cite{Xiang_LAMOSTDR5_2019} analysis of LAMOST DR5. For stars with [Fe/H] $>-1.5$, the H3 metallicities typically agree with the literature, but for more metal-poor stars they are higher than the SEGUE metallicities by $\sim0.2$ dex and lower than the \cite{Xiang_LAMOSTDR5_2019} values by $\sim0.1$ dex. Given that we found that the \cite{Xiang_LAMOSTDR5_2019} metallicities are possibly too high by $\sim 0.4$ dex (see Figure~\ref{fig:compare-feh}), both comparisons point to H3 having a slightly more metal-rich metallicity scale for very metal-poor stars. This can contribute to the discrepancy in the metallicity that we note for LMS-1 and Wukong.

\end{document}